\documentclass[prd,twocolumn,showpacs,nofootinbib]{revtex4}

\usepackage{amsfonts}

\begin{document}

\title{Corotating and irrotational binary black holes in quasi-circular orbits}

\author{Gregory B. Cook}
\email{cookgb@wfu.edu}
\affiliation{Department of Physics, Wake Forest University,
		 Winston-Salem, North Carolina 27109}

\date{\today}

\begin{abstract}
A complete formalism for constructing initial data representing
black-hole binaries in quasi-equilibrium is developed.  Radiation
reaction prohibits, in general, true equilibrium binary
configurations.  However, when the timescale for orbital decay is much
longer than the orbital period, a binary can be considered to be in
quasi-equilibrium.  If each black hole is assumed to be in
quasi-equilibrium, then a complete set of boundary conditions for {\em
all} initial data variables can be developed.  These boundary
conditions are applied on the apparent horizon of each black hole, and
in fact force a specified surface to be an apparent horizon.  A global
assumption of quasi-equilibrium is also used to fix some of the freely
specifiable pieces of the initial data and to uniquely fix the
asymptotic boundary conditions.  This formalism should allow for the
construction of completely general quasi-equilibrium black hole binary
initial data.
\end{abstract}


\maketitle

\section{Introduction}
\label{sec:introduction}
The problem of calculating the gravitational waveforms produced during
the final plunge and coalescence of a pair of black holes currently is
receiving considerable attention.  This attention is motivated by the
imminent deployment of the Laser Interferometric Gravitational Wave
Observatory (LIGO) and other, similar, detectors.  It is further
motivated by the fact that black-hole binaries are expected to be
strong sources of gravitational waves and are considered one of the
most likely candidates for early detection.  The detection and
interpretation of black hole mergers will be greatly facilitated by
theoretical predictions for the gravitational waveforms produced by
these events.  However, in the absence of predictions for the full
waveforms, even information about the behavior of the orbits will be
helpful.  In particular, information about the properties of the
so-called last stable orbit, or innermost stable circular orbit
(ISCO), is of interest.

Currently, large scale numerical simulations are the only known
approach for computing the gravitational waveforms produced during the
final plunge and coalescence of black-hole binaries.  Any dynamical
simulation must start with some initial data, and the results of such
a simulation are entirely dependent on how well the given initial data
correspond to the desired physical situation.  Thus, astrophysically
realistic initial data are required if we are going to simulate the
gravitational waveforms of astrophysical black-hole binaries.  Because
radiation damping tends to circularize binary orbits, orbits that are
nearly circular are of greatest interest.  Of course, energy loss by
gravitational radiation also means that no truly circular orbits can
exist for black-hole binary systems.  To keep this fact in mind, we
will always refer to such orbits as being quasi-circular.

The usefulness of current numerical simulations of general black-hole
binary systems is limited by two important facts.  First, the
simulations themselves are only beginning to reach the stage where
they can produce useful information.  Furthermore, the accuracy of
this information is still questionable.  Second, the initial data
currently being used for these simulations is rather crude.  Initial
data for binary black hole systems have been available for some time
now \cite{cook93,brandt_bruegmann97}.  The approaches for generating
these initial data can produce rather arbitrary configurations and a
way of picking which data sets represent binaries in quasi-circular
orbits was needed.  An approximation scheme for locating
quasi-circular orbits was developed\cite{cook94e}.  Based on an
effective-potential method, this scheme was first applied to
non-spinning, equal-mass binaries using the inversion-symmetric data
of Ref.~\cite{cook93}.  Later, this approach was applied to
non-spinning, equal-mass binaries based on puncture
data\cite{baumgarte-2000} and to spinning, equal-mass black
holes\cite{pfeiffer-etal-2000}.

Unfortunately, there are problems with these data sets.  Because the
schemes used to generate this data assume the metric to be conformally
flat, they cannot produce astrophysically accurate data for spinning
holes\cite{pfeiffer-etal-2000}.  Furthermore, the estimates for the
location of the ISCO produced by these schemes do not agree well with
results based on various post-Newtonian approximation
methods\cite{kidder92,Damour-etal-2000}.  There is some speculation
that this discrepancy is rooted in the approximation of conformal
flatness.  While the conformal flatness approximation certainly
introduces some error, I suspect that the most important source of
error comes from a failure to accurately approximate the masses of the
black holes as they approach the ISCO.

Initial-data schemes relying on the conformal flatness approximation
have been popular because they are relatively simple and inexpensive
from a computational point of view.  When maximal slicing is used, an
analytic solution of the momentum constraints
exists\cite{bowenyork80,ksy83}.  This means that only the Hamiltonian
constraint, a single quasi-linear elliptic PDE, need be solved.
Recently, efforts have been made to move beyond the approximation of
conformal flatness (cf.\ Ref.~\cite{Kerrbinary98,Kerrbinary2000}).
Unfortunately, these data have not yet been used to estimate the
location of the ISCO.

For the case of neutron-star binaries, a very different approach has
been used to construct initial data representing neutron stars in
quasi-circular orbits and to locate the ISCO.  At the core of this
approach is the assumption that the space-time admits an approximate
Killing vector, $\vec{\ell}\equiv(\partial/\partial{t}) +
\Omega(\partial/\partial\phi)$.  This approach was first tested for
the case of a single spinning neutron star\cite{cook96b}.
Subsequently, variations of this method were used to study
neutron-star binaries in quasi-circular
orbits\cite{baumgarte_etal97b,baumgarte_etal98b,baumgarte_etal98a,%
marronetti-etal-1998,bonazzola-etal-1999,marronetti-etal-1999,%
uryu-eriguchi-2000,uryu-etal-2000}.  This approach for generating
initial data for neutron-star binaries in quasi-circular orbits is
particularly attractive when compared to the effective potential
method used for black-hole binaries.  First, if the neutron-star
matter is to be in hydrostatic equilibrium, then the 4-velocity of the
matter must be proportional to the approximate Killing vector,
$\vec{\ell}$.  The assumption that $\vec{\ell}$ is a Killing vector
ultimately yields a relativistic Bernoulli equation that governs the
behavior of the matter and, most importantly, yields a unique value of
the angular velocity, $\Omega$, of the binary system in a circular
orbit at a given separation.  The presence of matter also sets a
natural length scale for the the problem and conservation of baryon
number makes it straightforward to connect initial-data solutions at
different radii to produce meaningful evolutionary sequences.

Recently, Gourgoulhon {\it et
al.}\cite{gourgoulhon-etal-2001a,gourgoulhon-etal-2001b} (hereafter
GGB1 and GGB2 respectively) have attempted to adapt the approach used
to construct neutron-star binary initial data to the case of
black-hole binaries.  Their initial results are very encouraging,
showing much better agreement with post-Newtonian techniques for
locating the ISCO.  However, there seem to be problems with their
solutions.  It seems likely that their initial data represent a much
better approximation to astrophysically realistic black-hole binaries
in quasi-circular orbit than the data produced using the effective
potential technique.  However, these new solutions do not represent
valid general relativistic initial data because they do not satisfy
the constraint equations everywhere.  Furthermore, as I will justify
later, I do not believe that the particular problem being solved in
GGB1,2 admits a solution.  Interestingly, this author had attempted to
solve virtually the identical problem\cite{cook94f} but was unable to
obtain convergence for the coupled system of equations.  As the
authors point out in GGB2, they must {\em regularize} their solution
for the shift in order to obtain an extrinsic curvature that does not
diverge at the throats of the black holes.  It is this regularization
technique that prohibits their solutions from satisfying the
constraints everywhere.  Although I have concerns regarding the
validity of the particular solutions found in GGB2, I believe that the
authors of GGB1 may have made a significant contribution to the effort
to construct astrophysically realistic black-hole binary initial data.
This contribution is rooted in their approach for determining the
orbital angular velocity, $\Omega$.

The problem with the approach taken in GGB1 is that it is not directly
applicable outside the restrictions of conformal flatness and maximal
slicing, and it seems that a solution within these limits does not
exist.  The limitations in the GGB1 method are rooted in the boundary
conditions posed on the inner boundaries of their solution domain,
that is, the black-hole boundaries.  In GGB1 the authors choose to
impose a particular isometry condition on the spatial hypersurface.
This isometry condition naturally yields boundary conditions on the
throats of the black holes, however, this isometry condition and the
boundary conditions derived from it have no direct connection to the
desired quasi-equilibrium state of the solution.  Interestingly, the
authors of GGB1 choose to mix the isometry-derived boundary conditions
with a Killing-horizon condition that is essentially an equilibrium
condition on the black hole horizon.  However, this mixed set of
boundary conditions is not consistent with the isometry condition
unless the artificial regularization of the shift is performed.
Again, this regularization prohibits the solutions from satisfying the
constraints everywhere.

In this paper, I will develop a formalism for constructing completely
general quasi-equilibrium black hole binary initial data.  This
formalism is also based on a generalization of the approach used for
constructing neutron-star binary initial data.  The key to this
formalism is a set of boundary conditions that are based on the
approximation that the black holes are in quasi-equilibrium.  Boundary
conditions for {\em all} of the quantities required to construct
quasi-equilibrium data can be found based on the quasi-equilibrium
assumption.  These boundary conditions are completely general.  In
particular, they do not depend on the often-used assumptions of
conformal flatness and maximal slicing.  This formalism is completed
by using the approach developed by GGB1 to determine the orbital
angular velocity $\Omega$ of the system, an approach that is also
rooted in the assumption of quasi-equilibrium.

In the following sections, I will derive and describe this new
formalism.  Where appropriate, I include comparisons with related
methods in order to clarify the advantages of the new formalism.  In
\S~\ref{sec:init-data-equat}, I begin with a general discussion of the
decomposition of Einstein's equations and how the initial-data
equations are posed.  In \S~\ref{sec:decomp-constr}, I discuss two
particular decompositions of the constraint equations.  In
\S~\ref{sec:bound-cond-constr}, I discuss the strengths and weaknesses
of the approach used in GGB1,2 and derive the boundary conditions
required for the new formalism.  In \S~\ref{sec:quasi-circ-orbits}, I
work through the remaining issues involved in constructing
quasi-equilibrium initial data. And in \S~\ref{sec:discussion}, I
conclude with discussion that focuses on the impact of possible
choices for the conformal 3-geometry and slicing condition.

\section{The initial-data equations}
\label{sec:init-data-equat}
In this work, we will use the standard $3+1$ decomposition with the
interval written as
\begin{equation}\label{eq:3+1-interval}
	{\rm d}s^2 = -\alpha^2{\rm d}t^2
		+ \gamma_{ij}({\rm d}x^i + \beta^i{\rm d}t)
				({\rm d}x^j + \beta^j{\rm d}t).
\end{equation}
where $\gamma_{ij}$ is the 3-metric induced on a $t=\text{const.}$
spatial hypersurface, $\alpha$ is the lapse function, and $\beta^i$
the shift vector.  The extrinsic curvature of the spatial slice,
$K_{ij}$, is defined by\footnote{In this paper, latin indices
denote 3-dimensional spatial indices.  Greek indices are 4-dimensional.
In \S~\ref{sec:quasi-equil-bound}, latin indices can be replaced with
indices varying over the two dimensions of the surface whenever only
2-dimensional tensors are involved in the expression.}
\begin{equation}\label{eq:extr-curv}
	K_{ij} \equiv -{\textstyle\frac12}{\cal L}_n\gamma_{ij},
\end{equation}
where ${\cal L}_n$ denotes the Lie derivative along the unit normal to
the spatial slice, $n^\mu$.  Einstein's equations, $G_{\mu\nu}~=~8\pi
T_{\mu\nu}$, then reduce to four sets of equations.  Two are evolution
equations for the spatial metric and extrinsic curvature:
\begin{equation}\label{eq:g-evolution}
	\partial_t \gamma_{ij} = -2\alpha K_{ij}
		+ \bar\nabla_i\beta_j
		+ \bar\nabla_j\beta_i
\end{equation}
and
\begin{eqnarray}\label{eq:K-evolution}
	\partial_t K_{ij} &=& \alpha \left[
		\bar{R}_{ij} - 2K_{i\ell}K^\ell_j + K K_{ij} \right.
		\\ && \mbox{\hspace{0.5in}} \left.
		- 8\pi S_{ij} + 4\pi \gamma_{ij}(S - \rho)\right]
		\nonumber \\ && \mbox{} 
		- \bar\nabla_i\bar\nabla_j\alpha
		+ \beta^\ell\bar\nabla_\ell K_{ij}
		+ K_{i\ell}\bar\nabla_j\beta^\ell
		+ K_{j\ell}\bar\nabla_i\beta^\ell. \nonumber
\end{eqnarray}
The remaining two are the constraint equations:
\begin{equation}\label{eq:Hamiltonian-const}
	\bar{R} + K^2 - K_{ij}K^{ij} = 16\pi \rho
\end{equation}
and
\begin{equation}\label{eq:momentum-const}
	\bar\nabla_j\left(K^{ij} - \gamma^{ij}K\right) = 8\pi j^i.
\end{equation}
Here, $\bar\nabla_i$, $\bar{R}_{ij}$, and $\bar{R}$ are, respectively,
the covariant derivative, Ricci tensor, and Ricci scalar associated
with the spatial metric $\gamma_{ij}$. $K \equiv K^i_i$, $S_{ij}$ is
the matter stress tensor, $S\equiv S^i_i$, $\rho$ is the matter energy
density, and $j^i$ is the matter momentum density.  These matter terms
are related to the stress-energy tensor via:
\begin{eqnarray}
\label{eq:S-def}
	S_{\mu\nu} &\equiv&
		\gamma_\mu^\alpha \gamma_\nu^\beta T_{\alpha\beta}, \\
\label{eq:j-def}
	j_\mu &\equiv& -\gamma_\mu^\nu n^\alpha T_{\nu\alpha}, \\
\label{eq:rho-def}
	\rho &\equiv& n^\mu n^\nu T_{\mu\nu} \\
\label{eq:T-decomp}
	T_{\mu\nu} &=& S_{\mu\nu} + 2n_{(\mu}j_{\nu)} + n_\mu n_\nu \rho.
\end{eqnarray}

Initial data for a Cauchy evolution via Einstein's equations consists
of specifying $\gamma_{ij}$ and $K_{ij}$ on an initial hypersurface.
The Hamiltonian constraint (\ref{eq:Hamiltonian-const}) and the
momentum constraints (\ref{eq:momentum-const}) constitute the
initial-data equations of general relativity, and the initial data
must satisfy these equations.  Equations~(\ref{eq:Hamiltonian-const})
and (\ref{eq:momentum-const}) {\em constrain} four of the 12 degrees of
freedom found in the initial data, $\gamma_{ij}$ and $K_{ij}$.  The
covariant nature of Einstein's equations allows us four {\em gauge}
degrees of freedom.  The remaining four degrees of freedom are the
{\em dynamical} degrees of freedom and are split two apiece between
the metric and extrinsic curvature.

The task of constructing initial data for a Cauchy evolution via
Einstein's equations can be broken down into a few fundamental
task shared by any approach.
\begin{enumerate}
\item A decomposition of the constraints must be chosen that specifies
how the constrained, gauge, and dynamical degrees of freedom are
associated with the initial data.
\item Boundary conditions must be chosen for the constrained degrees
of freedom such that a)~the constraint equations reduce to a
well-posed set of elliptic PDEs and b)~they are compatible with the
desired physical content of the initial data.
\item A choice for the spatial and temporal gauge must be made that is
compatible with the desired physical content of the initial data.
\item The dynamical degrees of freedom must be specified so as to yield
the desired physical content in the initial data.
\end{enumerate}
Once these steps have been completed, all that remains are the
technical aspects of solving the coupled set of elliptic PDEs.  The
{\em physics} lies entirely in the four enumerated steps given above.

\section{Decompositions of the constraints}
\label{sec:decomp-constr}
In the weak field limit where Einstein's equations can be linearized,
there are clear ways to determine which components of the initial
data are dynamic, which are constrained, and which are gauge.  However,
in the full nonlinear theory, there is no unique way to perform the
decomposition.  Perhaps the most widely used class of constraint
decompositions are the York--Lichnerowicz conformal decompositions
which are based on a conformal decomposition of the metric and various
other quantities\cite{lichnerowicz-1944,york-1971,york-1972}.

The metric is decomposed into a conformal factor, $\psi$, multiplying
a conformal metric:
\begin{equation}\label{eq:conformal-metric}
	\gamma_{ij} \equiv \psi^4\tilde\gamma_{ij}.
\end{equation}
Using (\ref{eq:conformal-metric}), the Hamiltonian
constraint (\ref{eq:Hamiltonian-const}) can be written as
\begin{equation}\label{eq:conf-Hamiltonian-const-v1}
	\tilde\nabla^2\psi - {\textstyle\frac18}\psi\tilde{R}
		- {\textstyle\frac18}\psi^5K^2
		+ {\textstyle\frac18}\psi^5K_{ij}K^{ij} = -2\pi \psi^5\rho,
\end{equation}
where $\tilde\nabla^2 \equiv \tilde\nabla^i\tilde\nabla_i$, and
$\tilde\nabla_i$ and $\tilde{R}$ are the covariant derivative and
Ricci scalar associated with $\tilde\gamma_{ij}$.  With this conformal
decomposition, the conformal metric, $\tilde\gamma_{ij}$ is considered
to encode the dynamical degrees of freedom of the metric {\em and} the
initial gauge choices for the three spatial coordinates.  The
conformal factor, $\psi$, is the constrained portion of the metric.

Most decompositions of the extrinsic curvature begin by splitting it
into its trace, $K$ and tracefree parts, $A_{ij}$,
\begin{equation}\label{eq:trace-free-K}
	K_{ij} \equiv A_{ij} + {\textstyle\frac13}\gamma_{ij}K.
\end{equation}
The decomposition proceeds by using the fact that we can covariantly
split any symmetric tracefree tensor, ${\cal S}^{ij}$, as
follows\cite{york-1973}:
\begin{equation}\label{eq:gen-TT-decomp}
	{\cal S}^{ij} \equiv ({\mathbb L}X)^{ij} + {\cal T}^{ij}.
\end{equation}
Here, ${\cal T}^{ij}$ is a symmetric, transverse-traceless tensor
(i.e., $\nabla_j{\cal T}^{ij} = 0$ and ${\cal T}^i_i=0$) and, in three
dimensions,
\begin{equation}\label{eq:TT-op-def}
	({\mathbb L}X)^{ij} \equiv \nabla^iX^j +\nabla^jX^i
		- {\textstyle\frac23}\gamma^{ij}\nabla_\ell X^\ell.
\end{equation}
At this point, the decomposition can proceed in several different
directions, and we will consider two of them below.  However, in all
cases, $K$ is considered to encode the initial temporal gauge choice,
${\cal T}^{ij}$ encodes the two dynamical degrees of freedom, and
$X^i$ are the three constrained degrees of freedom.

\subsection{Conformal transverse-traceless decomposition}
\label{sec:conf-transv-trac}
Historically, the most widely used decomposition expresses the
extrinsic curvature as
\begin{equation}\label{eq:decomp-K}
	K^{ij} \equiv \psi^{-10}\left[
		(\tilde{\mathbb L}X)^{ij} + \tilde{Q}^{ij}\right]
		+ {\textstyle\frac13}\gamma_{ij}K.
\end{equation}
In this case, the transverse-traceless decomposition takes place with
respect to the conformal metric.  It will also be convenient to define
a conformal trace-free extrinsic curvature, $\tilde{A}^{ij}$, by
\begin{equation}\label{eq:conf-A}
	\tilde{A}^{ij} \equiv \psi^{10}A^{ij} =
	(\tilde{\mathbb L}X)^{ij} + \tilde{Q}^{ij}.
\end{equation}
Using (\ref{eq:decomp-K}) in the momentum constraint
(\ref{eq:momentum-const}) yields
\begin{equation}\label{eq:conf-mom}
	\tilde\Delta_{\mathbb L}X^i =
		{\textstyle\frac23}\psi^6\tilde\nabla^iK
		+ 8\pi \psi^{10}j^i,
\end{equation}
where
\begin{eqnarray}\label{eq:long-elliptic-op}
	\tilde\Delta_{\mathbb L}X^i &\equiv&
		\tilde\nabla_j(\tilde{\mathbb L}X)^{ij} 
		\\ &=& \mbox{}
		\tilde\nabla^2X^i
		+ {\textstyle\frac13}\tilde\nabla^i(\tilde\nabla_jX^j)
		+ \tilde{R}^i_jX^j, \nonumber
\end{eqnarray}
and $\tilde{R}^i_j$ is the Ricci tensor associated with
$\tilde{\gamma}_{ij}$.  In terms of (\ref{eq:conf-A}), the Hamiltonian
constrain (\ref{eq:conf-Hamiltonian-const-v1}) becomes
\begin{eqnarray}\label{eq:conf-Hamiltonian-const-v2}
\tilde\nabla^2\psi - {\textstyle\frac18}\psi\tilde{R}
		- {\textstyle\frac1{12}}\psi^5K^2 & \\ \mbox{}
		+ {\textstyle\frac18}\psi^{-7}\tilde{A}_{ij}\tilde{A}^{ij}
		&=& -2\pi \psi^5\rho. \nonumber
\end{eqnarray}
Together, Eqns.~(\ref{eq:conformal-metric}) and
(\ref{eq:decomp-K})--(\ref{eq:conf-Hamiltonian-const-v2}) constitute
the full conformal transverse-traceless decomposition of the
constraints.  To solve this set of equations, one must specify
$\tilde\gamma_{ij}$, $\tilde{Q}^{ij}$, and $K$ in addition to the
matter terms $\rho$ and $j^i$.

The reason that this decomposition has been so widely used stems from
the simplifications that occur when the freely specifiable data are
chosen so that the initial data hypersurface is maximal, $K=0$,
conformally flat so that $\tilde\gamma_{ij}$ is a flat metric, and
$\tilde{Q}^{ij}$ is chosen to vanish.  Then, for vacuum spacetimes,
the constraints reduce to
\begin{equation}\label{eq:conf-Ham-simp}
\tilde\nabla^2\psi
	 + {\textstyle\frac18}\psi^{-7}\tilde{A}_{ij}\tilde{A}^{ij} = 0,
\end{equation}
and
\begin{equation}\label{eq:conf-mom-simp}
	\tilde\nabla^2X^i
		+ {\textstyle\frac13}\tilde\nabla^i(\tilde\nabla_jX^j) = 0,
\end{equation}
and $\tilde\nabla_i$ is, in this case, the usual flat space covariant
derivative operator.  What is remarkable about this simplified system
of equations is that Eq.~(\ref{eq:conf-mom-simp}) decouples completely
from (\ref{eq:conf-Ham-simp}) and has analytic solutions that
represent a single black hole with any desired value of linear and
angular momentum\cite{bowen79,bowenyork80}.  Furthermore, since
(\ref{eq:conf-mom-simp}) is linear, any number of these solutions may
be superposed to represent multiple black holes.  All that is required
to construct complete initial data is to solve the single quasi-linear
elliptic PDE in Eq.~(\ref{eq:conf-Ham-simp}).

All of the work to date that has used the effective potential
method to locate quasi-circular orbits in black-hole binary
systems\cite{cook94e,baumgarte-2000,pfeiffer-etal-2000} has used
initial data computed by this system.  The only difference between
the data used in Refs.~\cite{cook94e,pfeiffer-etal-2000}, and in
Ref.~\cite{baumgarte-2000} is in the topology of the initial data
hypersurfaces.

\subsection{Conformal thin-sandwich decomposition}
\label{sec:conf-thin-sandw}
An alternative decomposition recently proposed by York\cite{york-1999}
is a generalization of an approach used first by
Wilson\cite{Frontiers:Wilson} and later by
others\cite{cook96b,baumgarte_etal97b,bonazzola97} for the study of
neutron-star binaries.  Furthermore, aside from some trivial
differences in conformal scalings, it also subsumes the approach
outlined in GGB1.  This decomposition differs from the conformal
transverse-traceless decomposition significantly in philosophy, but is
very similar in form.  The difference is found in the decomposition of
the extrinsic curvature.  In the conformal thin-sandwich
decomposition, $K_{ij}$ is decomposed as
\begin{equation}\label{eq:TS-decomp-K}
	K^{ij} = \frac{\psi^{-10}}{2\tilde\alpha}\left(
		(\tilde{\mathbb L}\beta)^{ij} - \tilde{u}^{ij}\right)
		+ {\textstyle\frac13}\gamma_{ij}K,
\end{equation}
or 
\begin{equation}\label{eq:TS-decomp-A}
	\tilde{A}^{ij} = \frac1{2\tilde\alpha}\left(
		(\tilde{\mathbb L}\beta)^{ij} - \tilde{u}^{ij}\right).
\end{equation}
Here, $\tilde{u}^{ij}$ takes the place of $\tilde{Q}^{ij}$ in
representing the freely specifiable dynamical portion of the extrinsic
curvature.  However, owing to the nature of the thin sandwich
decomposition, it also has a more physical meaning as we shall see in
a moment.  Notice that in Eq.~(\ref{eq:TS-decomp-A}), it is not the
conformal trace-free extrinsic curvature that is decomposed, but
rather $2\tilde\alpha\tilde{A}^{ij}$.  The new scalar, $\tilde\alpha$,
is the conformal lapse defined by
\begin{equation}\label{eq:conformal-lapse}
	\alpha \equiv \psi^6\tilde\alpha.
\end{equation}
Also notice that, in the transverse-traceless decomposition, the vector
$X^i$ has been replaced by the {\em shift vector}, $\beta^i$.

Using the thin-sandwich decomposition (\ref{eq:TS-decomp-K}) in the
momentum constraint (\ref{eq:momentum-const}) yields
\begin{eqnarray}\label{eq:TS-momentum-const}
	\tilde\Delta_{\mathbb L}\beta^i
		- (\tilde{\mathbb L}\beta)^{ij}
				\tilde\nabla_{\!j}\!\ln\tilde\alpha
		&=& {\textstyle\frac43}\tilde\alpha\psi^6\tilde\nabla^iK
		\\ \mbox{} & &
		+\,\tilde\alpha\tilde\nabla_{\!j}\!\left(
			{\textstyle\frac1{\tilde\alpha}}\tilde{u}^{ij}\right)
		+ 16\pi \tilde\alpha\psi^{10}j^i. \nonumber
\end{eqnarray}
Taken together, Eqns.~(\ref{eq:conformal-metric}),
(\ref{eq:TS-decomp-K}), (\ref{eq:TS-decomp-A}),
(\ref{eq:TS-momentum-const}), (\ref{eq:long-elliptic-op}), and
(\ref{eq:conf-Hamiltonian-const-v2}) constitute the full thin-sandwich
decomposition of the constraints.  To solve this set of equations, one
must specify $\tilde\gamma_{ij}$, $\tilde{u}^{ij}$, $K$, {\em and}
$\tilde\alpha$ in addition to the matter terms $\rho$ and $j^i$.

The thin-sandwich decomposition differs from most initial-data
decompositions in that its derivation uses a second spatial
hypersurface that is infinitesimally separated from the primary
spatial hypersurface.  The time and spatial coordinates in these two
hypersurfaces are connected by the kinematical variables, $\alpha$ and
$\beta^i$, and this connection allows us to obtain a physical
interpretation for $\tilde{u}^{ij}$, the freely specifiable dynamical
portion of the extrinsic curvature.  The result is that
$\tilde{u}^{ij}$ is a symmetric, tracefree tensor with the physical
interpretation that
\begin{equation}\label{eq:TS-u-def}
	\tilde{u}_{ij} \equiv \partial_t\tilde\gamma_{ij},
\end{equation}
and $t$ is the time coordinate associated with the time-like vector
$t^\mu$ that connects points in the two adjacent hypersurfaces that
carry the same spatial coordinate labels, i.e.
\begin{equation}\label{eq:t-def}
	t^\mu \equiv \alpha n^\mu + \beta^\mu,
\end{equation}
and $n^\mu$ is the timelike unit normal to the primary spatial
hypersurface.

This decomposition has proven useful for constructing
quasi-equilibrium initial data sets because, in the frame where the
configuration is assumed to be time-independent (the {\em corotating
frame}), it is a natural approximation to set $\tilde{u}_{ij}=0$.  To
my knowledge, all of the work so far (including
Refs.~\cite{Frontiers:Wilson,cook96b,baumgarte_etal97b,%
marronetti-etal-1998,bonazzola97}, and GGB1,2) use the same set of
assumptions to fix the freely specifiable data in the thin-sandwich
decomposition.  In particular, they all choose the initial data
hypersurface to be maximal, $K=0$, conformally flat so that
$\gamma_{ij}$ is a flat metric, $\tilde{u}_{ij}$ is chosen to be zero,
and they all fix the lapse by demanding that $\partial_tK=0$.  The
final assumption is a reasonable one since a quasi-equilibrium
solution was the goal of each of these works and the time vector
(\ref{eq:t-def}) is the one corresponding to the corotating frame.

This conformal thin-sandwich decomposition will form the foundation of
the initial-data formalism being developed.  However, unlike previous
works, we will only make the restriction that $\tilde{u}_{ij}=0$.  In
particular, we will {\em not} restrict the formalism to conformally
flat maximal slices of the full spacetime, and we {\em will} require
that all boundary conditions and any further restrictions apply (at
least in principle) to general slicings and conformal 3-geometries.

\section{Boundary conditions for the constraint equations}
\label{sec:bound-cond-constr}
\subsection{The GGB approach}
\label{sec:ggb-approach}
We now return to the specific problem of using the thin-sandwich
decomposition to construct initial data for a pair of black holes
in quasi-circular orbits.  In addition to the standard assumptions
outlined at the end of \S~\ref{sec:conf-thin-sandw}, GGB1 also
makes the assumption that the initial-data hypersurface is inversion
symmetric.  That is, the initial data hypersurface is taken to consist
of two asymptotically flat hypersurfaces that are joined together
at the throats of each black hole, and are isometric to each other
with each throat being a fixed point set of the isometry condition.
The isometry condition then imposes a set of boundary conditions
at the throats on all quantities on the manifold.  There are actually
two choices for the boundary conditions, and GGB1 make the choice
that requires the lapse to vanish on the throats.  For the shift
vector, the isometry imposes the following boundary conditions
\begin{eqnarray}
\label{eq:shift-isom-BC-1}
	\beta^r|_{\cal S} &=& 0, \\
\label{eq:shift-isom-BC-2}
	\left.\frac{\partial\beta^\theta}{\partial_r}\right|_{\cal S} &=& 0, \\
\label{eq:shift-isom-BC-3}
	\left.\frac{\partial\beta^\phi}{\partial_r}\right|_{\cal S} &=& 0,
\end{eqnarray}
where we are assuming the throat is a coordinate sphere and $|_{\cal
S}$ denotes evaluation on the throat.

Since this version of inversion symmetry requires that $\alpha|_{\cal
S}=0$, regularity of the extrinsic curvature at the throats requires
\begin{equation}\label{eq:GGB-regularity}
	(\tilde{\mathbb L}\beta)^{ij}|_{\cal S}=0.
\end{equation}
As GGB1 point out, this condition is potentially problematical.

Given the choice for the isometry condition used in GGB1, it
follows\cite{cook90} that the throats are apparent horizons.  GGB1
make the interesting choice of using this fact to change boundary
conditions (\ref{eq:shift-isom-BC-2}) and (\ref{eq:shift-isom-BC-3})
on the shift.  They obtain new boundary conditions on the shift
by demanding that $\partial/\partial_t$ be null on the throats so
that the throats are Killing horizons.  For the case that the
lapse vanishes on the throats, this condition yields
\begin{equation}\label{eq:GGB-shift-BC}
	\beta^i|_{\cal S}=0.
\end{equation}
Altering the boundary conditions on the shift in this way poses a
problem.  Using Eq.~(\ref{eq:GGB-shift-BC}) to replace conditions
(\ref{eq:shift-isom-BC-2}) and (\ref{eq:shift-isom-BC-3}) means that
the solution is not guaranteed to be inversion-symmetric.  This, in
turn, means that the inner boundaries are not guaranteed to be
apparent horizons, from which it follows the Killing horizon boundary
conditions may not be appropriate for the inner boundaries.

In GGB1, the authors go to great length to show that
inversion-symmetry, their Killing-horizon boundary conditions on the
shift, and the required regularity condition (\ref{eq:GGB-regularity})
can be compatible.  However, satisfying the Killing-horizon boundary
conditions together with {\em some} of the inversion-symmetry boundary
conditions does not guarantee that the regularity condition
(\ref{eq:GGB-regularity}) is satisfied, or that the resulting solution
is inversion symmetric.  If fact, as pointed out in GGB2, it seems
that their solutions are not compatible with this condition without
some kind of artificial regularization of their solution for the
shift.  This is not surprising.  We can rewrite the evolution equation
for the metric as
\begin{eqnarray}\label{eq:TS-gdot-decomp}
	\partial_t\gamma_{ij} &=& u_{ij}
		+ {\textstyle\frac23}\gamma_{ij}\left(
			\bar\nabla_k\beta^k - \alpha K\right) \\
		 &=& \psi^4\left[\tilde{u}_{ij}
		+ {\textstyle\frac23}\tilde\gamma_{ij}\left(
			\tilde\nabla_k\beta^k + 6\beta^k\tilde\nabla_k\ln\psi
			- \psi^6\tilde\alpha K \right)\right], \nonumber
\end{eqnarray}
with $u_{ij}\equiv\psi^4\tilde{u}_{ij}$.  Given the the
Killing-horizon boundary conditions and the choices of maximal slicing
and $\tilde{u}_{ij}=0$, Eq.~(\ref{eq:TS-gdot-decomp}) reduces to
\begin{equation}\label{eq:TS-GGB-gdot-behavior}
	\partial_t\gamma_{ij}|_{\cal S} =
		{\textstyle\frac23}\psi^4\tilde{\gamma}_{ij}\left.
			\frac{\partial\beta^r}{\partial_r}\right|_{\cal S}.
\end{equation}
It is demonstrated in GGB1 that regularity requires
$\partial\beta^r/\partial_r|_{\cal S}=0$ and the authors take
Eq.~(\ref{eq:TS-GGB-gdot-behavior}) as establishing regularity.
However, the choice $\tilde{u}_{ij}=0$ is {\em not} equivalent to
$\partial_t\gamma_{ij}=0$, but only fixes the time rate of change of
the conformal metric.  The time rate of change of the conformal factor
is not fixed by this condition and is only guaranteed to vanish if
$\vec{t}$ is a true Killing vector.  Thus, for a quasi-equilibrium
solution of the constraints, it seems reasonable that
$\partial\beta^r/\partial_r|_{\cal S}$ is small in some appropriate
norm, but it is unlikely to vanish.

So, the approach outlined in GGB1 has two technical problems: the
boundary conditions being used do not guarantee that the solution is
inversion symmetric or that the solution will be regular at the inner
boundary.  Both of these problems are fixed in GGB2 by an artificial
regularization of the shift solution.  The unfortunate side effect of
this regularization procedure is that the solutions no longer satisfy
the constraint equations and, thus, cannot represent valid initial
data.  In fact, it seems that a regular, inversion-symmetric solution
for the extrinsic curvature is not possible (or is at least extremely
unlikely), given the assumptions of GGB1.

\subsection{Killing-horizon boundary conditions}
\label{sec:kill-horiz-bound}
Although the particular set of boundary conditions proposed in GGB1
does not seem to yield a well-posed system of elliptic equations, the
idea of using some kind of boundary conditions on the apparent
horizons of the black holes is promising.  An alternate approach for
obtaining boundary conditions on the shift is found by a slight
generalization of the the Killing-horizon conditions proposed in GGB1.
As we will see, this change in perspective will allow us to derive
boundary conditions on the shift that will yield corotating black
holes, irrotational black holes, or black holes with some arbitrary
angular velocity.  We begin by noting that the outward pointing null
vector whose vanishing expansion defines the apparent horizon can be
written as
\begin{equation}\label{eq:AH-null-vec-def}
	k^\mu \propto \left(n^\mu + s^\mu\right).
\end{equation}
Here, $s^\mu$ is the outward pointing unit vector normal to the
apparent horizon (see \S~\ref{sec:quasi-equil-bound} for further
details).  In terms of components, this can be written as
\begin{equation}\label{eq:AH-null-vec-comp}
	k^\mu = \left[1,\alpha s^i - \beta^i\right],
\end{equation}
where I have chosen a normalization so that $k^t\equiv1$.  Note that
$k^\mu$ is null by construction and $k^\mu k_\mu=0$ for {\em any}
choice of the shift vector.

If we assume rotation in the $\phi$ direction, then the approximate
killing vector for a quasi-equilibrium configuration is given by
\begin{equation}\label{eq:corot-approx-kv}
	\vec\ell \equiv \frac\partial{\partial t}
		+ \Omega\frac\partial{\partial \phi},
\end{equation}
where $\Omega$ is the orbital angular velocity of the binary, and
$\partial/\partial\phi$ represent the asymptotic Killing vector for
rotation in the $\phi$ direction.  In the corotating coordinate
system, $\vec\ell$ is the time vector and it has components
$\ell^\mu=[1,0,0,0]$.  If the individual black holes are corotating
with the system, then they do not appear to be rotating in that
coordinate system.  In this case, the Killing-horizon condition
implies that $\ell^\mu$ should be parallel to $k^\mu$ which
immediately yields
\begin{equation}\label{eq:corot-shift-bc}
	\beta^i|_{\cal S} = \alpha s^i|_{\cal S}.
\end{equation}
Notice that this condition reduces directly to
Eq.~(\ref{eq:GGB-shift-BC}) when $\alpha|_{\cal S}=0$.

We can also derive a boundary condition for black holes that are
not rotating with respect to observers at rest at infinity, black holes
that we might loosely refer to as being irrotational.  In this case,
the outward pointing null vectors at the apparent horizon should
be parallel to $\vec{t}$, not to $\vec\ell$.  In the corotating
coordinate system, the components of $\vec{t}$ can be written 
symbolically as
\begin{equation}\label{eq:corot-t-coords}
	t^\mu = [1,-\Omega(\partial/\partial\phi)^i].
\end{equation}
It follows immediately that the appropriate boundary conditions for
irrotational black holes are
\begin{equation}\label{eq:irrot-shift-bc-1}
	\beta^i|_{\cal S} = \alpha s^i|_{\cal S}
		+ \Omega\left.\left(\frac\partial{\partial\phi}\right)^{\!\!i}
		\right|_{\cal S}.
\end{equation}

For both boundary conditions, corotation given by
(\ref{eq:corot-shift-bc}), and irrotational given by
(\ref{eq:irrot-shift-bc-1}), the conditions are given in terms of the
corotating coordinate system.  This means that the appropriate
asymptotic boundary conditions for the shift are that
\begin{equation}\label{eq:asympt-shift-bc}
	\beta^i|_{{\rm r}\to\infty} 
		\sim \Omega\left(\frac\partial{\partial\phi}\right)^{\!\!i}.
\end{equation}

The corotation and irrotational conditions can be generalized to
yield a boundary condition for an equilibrium black holes with
arbitrary spin.  Let $\chi^i$ denote a flat-space Killing vector
for rotation in an arbitrary direction, just as $(\partial/\partial\phi)^i$
represents rotation about in the $\phi$ direction.  With $\Omega_\chi$
representing the angular velocity of rotation of the black holes as
measured by observers at rest at infinity, we find that the components
of the null vectors generating the horizon are
\begin{equation}\label{eq:arb-rot-t-coord}
	[1,\Omega_\chi\chi^i-\Omega(\partial/\partial\phi)^i],
\end{equation}
when written in terms of the corotating coordinate system.  This
immediately yields the boundary conditions for a quasi-equilibrium
black hole with arbitrary spin:
\begin{equation}\label{eq:arb-rot-shift-bc-1}
	\beta^i|_{\cal S} = \alpha s^i|_{\cal S} - \Omega_\chi\chi^i|_{\cal S}
		+ \Omega\left.\left(\frac\partial{\partial\phi}\right)^{\!\!i}
		\right|_{\cal S}.
\end{equation}
As we will see below, we will need to be a bit more careful in
defining the shift boundary condition for the irrotational and general
spin cases, but (\ref{eq:irrot-shift-bc-1}) and
(\ref{eq:arb-rot-shift-bc-1}) give us a good starting point for these
conditions.

\subsection{Quasi-equilibrium boundary conditions}
\label{sec:quasi-equil-bound}
The choice of the particular inversion-symmetric class of black-hole
initial data used in GGB1 was made because it guarantees that the
inner boundary surfaces will be apparent horizons.  If we are to give
up inversion-symmetry, we must find some other way of fixing boundary
conditions at the inner-boundary surfaces for all of the quantities
needed to construct initial data.  These boundary conditions can be
found by forcing the inner boundary to be an apparent horizon and by
assuming that the associated black hole is in quasi-equilibrium.
Pieces of the resulting approach have been worked out by
Thornburg\cite{thornburg87} and by Eardley\cite{eardley-1998}, but to
my knowledge have not been adapted for constructing quasi-equilibrium
initial data.  Because there are several errors in the derivations and
results in Ref.~\cite{eardley-1998}, and to provide a uniform
notation, I will rederive all of the equations here.

Our inner boundary surface, ${\cal S}$, is assumed to be a spacelike
2-surface with topology $S^2$.  Because ${\cal S}$ is closed, we can
define its unit normal via
\begin{equation}\label{eq:AH-normal-parm}
	s_i \equiv \lambda \bar\nabla_i\tau,
\end{equation}
where $\lambda$ is a normalization constant fixed by $s^is_i\equiv1$
and $\tau$ is a scalar function which has a level surface $\tau=0$
that defines the surface ${\cal S}$.  The 4-dimensional generalization
of $s^i$ has components $s^\mu = [0,s^i]$ obtained from the condition
that $s^\mu n_\mu=0$.

The metric, $h_{ij}$, induced on ${\cal S}$ by $\gamma_{ij}$ is
given by
\begin{equation}\label{eq:2-metric-def}
	h_{ij} \equiv \gamma_{ij} - s_i s_j.
\end{equation}
We also define the extrinsic curvature, $H_{ij}$, of ${\cal S}$
embedded in the 3-dimensional spatial hypersurface as
\begin{equation}\label{eq:2-EC-def}
	H_{ij} \equiv -h_i^k h_j^\ell \bar\nabla_{(k}s_{\ell)}
		= -{\textstyle\frac12}{\cal L}_s h_{ij}.
\end{equation}

Naturally associated with ${\cal S}$ are two sets of null vectors: a
set of outgoing null rays, $k^\mu$, and a set if ingoing null rays,
$\acute{k}^\mu$, defined by
\begin{equation}\label{eq:null-vec-defs}
	k^\mu = {\textstyle\frac1{\sqrt2}}\left(n^\mu + s^\mu\right)
	\quad\text{and}\quad
	\acute{k}^\mu = {\textstyle\frac1{\sqrt2}}\left(n^\mu - s^\mu\right).
\end{equation}
Associated with each set of null rays is an extrinsic curvature of 
${\cal S}$ as embedded in the full 4-dimensional manifold.  These
are defined as
\begin{eqnarray}
\label{eq:null-ec-def1}
	\Sigma_{\mu\nu} &\equiv&
		-{\textstyle\frac12}h_\mu^\alpha h_\nu^\beta
				{\cal L}_k g_{\alpha\beta}, \\
\label{eq:null-ec-def2}
	\acute\Sigma_{\mu\nu} &\equiv&
		-{\textstyle\frac12}h_\mu^\alpha h_\nu^\beta
				{\cal L}_{\acute{k}} g_{\alpha\beta}.
\end{eqnarray}
To simplify the following, we will introduce various projections of
$K_{ij}$ along and normal to ${\cal S}$:
\begin{eqnarray}
\label{eq:J2-def}
	J_{ij} &\equiv& h_i^k h_j^\ell K_{k\ell}, \\
\label{eq:J1-def}
	J_i &\equiv& h_i^k s^\ell K_{k\ell}, \\
\label{eq:J0-def}
	J &\equiv& h^{ij}J_{ij} = h^{ij}K_{ij}, \\
\label{eq:K-via-J}
	K_{ij} &=& J_{ij} + 2s_{(i}J_{j)} + s_i s_j(K-J).
\end{eqnarray}
We can then simplify Eqns.~(\ref{eq:null-ec-def1}) and
(\ref{eq:null-ec-def2}) to
\begin{equation}\label{eq:null-ec-defs3}
	\Sigma_{ij} = {\textstyle\frac1{\sqrt2}}\left(J_{ij} + H_{ij}\right)
	\quad\text{and}\quad
	\acute\Sigma_{ij} = 
		{\textstyle\frac1{\sqrt2}}\left(J_{ij} - H_{ij}\right)
\end{equation}
Now, we define the expansion of outgoing null rays, $\sigma$, and
ingoing null rays, $\acute\sigma$, via
\begin{eqnarray}
\label{eq:null-expansion-def1}
	\sigma &\equiv& h^{ij}\Sigma_{ij} = 
		{\textstyle\frac1{\sqrt2}}\left(J + H\right), \\
\label{eq:null-expansion-def2}
	\acute\sigma &\equiv& h^{ij}\acute\Sigma_{ij} = 
		{\textstyle\frac1{\sqrt2}}\left(J - H\right).
\end{eqnarray}
Finally, we will define the shear of the outgoing null rays,
$\sigma_{ij}$, and ingoing null rays, $\acute\sigma_{ij}$, via
\begin{equation}\label{eq:shear-defs}
	\sigma_{ij} \equiv \Sigma_{ij} - {\textstyle\frac12}h_{ij}\sigma
	\quad\text{and}\quad
	\acute\sigma_{ij} \equiv \acute\Sigma_{ij}
		- {\textstyle\frac12}h_{ij}\acute\sigma.
\end{equation}

In order to generate boundary conditions, we need to make some
assumptions.  First, we will demand that our inner boundary, ${\cal
S}$, is an apparent horizon.  This is equivalent to demanding that the
expansion of the outgoing null rays vanishes on ${\cal S}$.  Thus our
first condition is that
\begin{equation}\label{eq:AH-cond}
	\sigma = 0.
\end{equation}
If we are looking for quasi-equilibrium configurations, then we want
${\cal S}$ to remain at the same coordinate location in the
3-dimensional hypersurface as time passes.  To enforce this condition,
it is necessary that ${\cal L}_t\tau = \partial_t\tau=0$, where
$\vec{t}$ is the approximate Killing vector associated with our demand
of quasi-equilibrium.  However, there is no reason that coordinates
cannot be free to move {\em on} ${\cal S}$.  In fact, this freedom is
necessary to allow for rotation of the black hole.  This means that,
in order to keep ${\cal S}$ at the same coordinate location,
$\partial_t\tau=0$ is too strong a condition.  If we define
$\vec\zeta$ as the part of $\vec{t}$ that is orthogonal to ${\cal S}$,
then it is necessary and sufficient that
\begin{equation}\label{eq:AH-fix-cond}
	{\cal L}_\zeta\tau=0 \quad\text{and}\quad
	h_i^j\bar\nabla_j{\cal L}_\zeta\tau = \hat\nabla_i{\cal L}_\zeta\tau=0,
\end{equation}
where $\hat\nabla_i$ is the 2-dimensional covariant derivative
compatible with $h_{ij}$.  If we define the normal component of the
shift as
\begin{equation}\label{eq:beta-perp-def}
	\beta_\perp \equiv \beta^is_i,
\end{equation}
then we can write $\vec\zeta$ as
\begin{equation}\label{eq:l-def}
	\zeta^\mu \equiv \alpha n^\mu + \beta_\perp s^\mu,
\end{equation}

We now need to consider how the expansions, $\sigma$ and
$\acute\sigma$, evolve along $\vec\zeta$.  For this calculation, a few
identities are crucial:
\begin{eqnarray}
\label{eq:ident-1}
	h^{ij}\bar{R}_{ij} &=& {\textstyle\frac12}(
		\bar{R} - H^2 + H_{ij}H^{ij} + \hat{R}), \\
\label{eq:ident-2}
	s^i s^j\bar{R}_{ij} &=& {\textstyle\frac12}(
		\bar{R} + H^2 - H_{ij}H^{ij} - \hat{R}), \\
\label{eq:ident-3}
	\lambda\left[{\cal L}_\zeta,\bar\nabla_\mu\right]\tau &=&
		-n_\mu\left[n^\nu\nabla_\nu\beta_\perp +
			\beta_\perp(K-J)\right],
\end{eqnarray}
where $\hat{R}$ is the Ricci scalar associated with $h_{ij}$ and
$\nabla_\mu$ is the 4-dimensional covariant derivative.  The first two
identities are obtained from various combinations and contractions of
the Gauss--Codazzi--Ricci equations that govern the embedding of
${\cal S}$ in the the 3-dimensional hypersurface.  The third
identity can be obtained by direct calculation.

As useful intermediate results, we obtain:
\begin{eqnarray}
\label{eq:J-dot}
	{\cal L}_\zeta J &=&
		\Bigl[{\textstyle\frac12}(J^2 - H^2)
			+ {\textstyle\frac12}(J_{ij}J^{ij} + H_{ij}H^{ij})
	\nonumber \\ \mbox{}&& \hspace{0.5in}
			- J_iJ^i + 8\pi S_{ij}s^is^j
			+ {\textstyle\frac12}\hat{R}
			- \hat\nabla^2\Bigr]\alpha
	\nonumber \\ \mbox{}&&
		+ \Bigl[J_{ij}H^{ij} + JH - HK - 8\pi j_i s^i
	\nonumber \\ \mbox{}&& \hspace{0.5in}
			+ \hat\nabla_iJ^i + 2J^i\hat\nabla_i\Bigr]\beta_\perp
		+ H s^i\bar\nabla_i\alpha, \\
\label{eq:H-dot}
	{\cal L}_\zeta H &=&
		\left[J_{ij}H^{ij} - 8\pi j_i s^i - \hat\nabla_iJ^i
			- 2J^i\hat\nabla_i\right]\alpha
	\nonumber \\ \mbox{}&&
		+ \Bigl[{\textstyle\frac12}(J^2 + H^2)
			+ {\textstyle\frac12}(J_{ij}J^{ij} + H_{ij}H^{ij})
	\nonumber \\ \mbox{}&& \hspace{0.5in}
			+ J_iJ^i
			- JK
			+ 8\pi\rho
			- {\textstyle\frac12}\hat{R}
			+ \hat\nabla^2\Bigr]\beta_\perp
	\nonumber \\ \mbox{}&&
		+ J s^i\bar\nabla_i\alpha.
\end{eqnarray}
These equations follow from the identities
(\ref{eq:ident-1})--(\ref{eq:ident-3}) and the evolution and
constraint equations
(\ref{eq:g-evolution})--(\ref{eq:momentum-const}).  Finally, we can
simplify these to the desired results:
\begin{eqnarray}
\label{eq:og-exp-dot}
	{\cal L}_\zeta\sigma &=& {\textstyle\frac1{\sqrt2}}\left[
		\sigma(\sigma + {\textstyle\frac12}\acute\sigma
		- {\textstyle\frac1{\sqrt2}}K) + {\cal E}\right]
			(\beta_\perp + \alpha)
	\nonumber \\ \mbox{}&&
		{\textstyle\frac1{\sqrt2}}\Bigl[
		\sigma({\textstyle\frac12}\sigma
		- {\textstyle\frac12}\acute\sigma
		- {\textstyle\frac1{\sqrt2}}K)
		+ {\cal D}
	\nonumber \\ \mbox{}&& \hspace{0.3in}
		+ 8\pi T_{\mu\nu}k^\mu\acute{k}^\nu\Bigr]
			(\beta_\perp - \alpha)
		+ \sigma s^i\bar\nabla_i\alpha, \\
\label{eq:ig-exp-dot}
	{\cal L}_\zeta\acute\sigma &=& -{\textstyle\frac1{\sqrt2}}\left[
		\acute\sigma(\acute\sigma + {\textstyle\frac12}\sigma
		- {\textstyle\frac1{\sqrt2}}K) + \acute{\cal E}\right]
			(\beta_\perp - \alpha)
	\nonumber \\ \mbox{}&&
		- {\textstyle\frac1{\sqrt2}}\Bigl[
		\acute\sigma({\textstyle\frac12}\acute\sigma
		- {\textstyle\frac12}\sigma
		- {\textstyle\frac1{\sqrt2}}K)
		+ \acute{\cal D}
	\nonumber \\ \mbox{}&& \hspace{0.3in}
		+ 8\pi T_{\mu\nu}k^\mu\acute{k}^\nu\Bigr]
			(\beta_\perp + \alpha)
		- \acute\sigma s^i\bar\nabla_i\alpha,
\end{eqnarray}
where
\begin{eqnarray}
\label{eq:D1-def}
	{\cal D} &\equiv& h^{ij}(\hat\nabla_i + J_i)(\hat\nabla_j + J_j)
		- {\textstyle\frac12}\hat{R}, \\
\label{eq:D2-def}
	\acute{\cal D} &\equiv& h^{ij}(\hat\nabla_i - J_i)(\hat\nabla_j - J_j)
		- {\textstyle\frac12}\hat{R}, \\
\label{eq:E1-def}
	{\cal E} &\equiv& \sigma_{ij}\sigma^{ij}
		+ 8\pi T_{\mu\nu}k^\mu k^\nu, \\
\label{eq:E2-def}
	\acute{\cal E} &\equiv& \acute\sigma_{ij}\acute\sigma^{ij}
		+ 8\pi T_{\mu\nu}\acute{k}^\mu \acute{k}^\nu.
\end{eqnarray}
We note that ${\cal E}$ and $\acute{\cal E}$ are both non-negative so
long as the matter satisfies either the weak or the strong energy
condition.  The two equations for the evolution of the expansions,
(\ref{eq:og-exp-dot}) and (\ref{eq:ig-exp-dot}), are not yet
restricted by any assumption of quasi-equilibrium.  The only
assumption in their derivation is that the surface ${\cal S}$ remain
at a constant coordinate location which is embodied in
(\ref{eq:AH-fix-cond}).

If we now restrict Eqns.~(\ref{eq:og-exp-dot}) and (\ref{eq:ig-exp-dot})
to apply to apparent horizons by enforcing condition (\ref{eq:AH-cond})
and also impose quasi-equilibrium on the evolution equations for the
expansions,
\begin{equation}\label{eq:QE-expans}
	{\cal L}_\zeta\sigma = 0 \quad\text{and}\quad
	{\cal L}_\zeta\acute\sigma = 0,
\end{equation}
then we find
\begin{eqnarray}
\label{eq:QE-og-exp1}
	 0 &=& {\cal D}(\beta_\perp - \alpha), \\
\label{eq:QE-ig-exp1}
	\acute\sigma s^i\bar\nabla_i\alpha &=&
	- {\textstyle\frac1{\sqrt2}}\left[
		\acute\sigma(\acute\sigma
		- {\textstyle\frac1{\sqrt2}}K)
		+ \acute\sigma_{ij}\acute\sigma^{ij}\right]
			(\beta_\perp - \alpha)
	\nonumber \\ \mbox{}&&
		- {\textstyle\frac1{\sqrt2}}\Bigl[
		\acute\sigma({\textstyle\frac12}\acute\sigma
		- {\textstyle\frac1{\sqrt2}}K)
		+ \acute{\cal D}\Bigr]
			(\beta_\perp + \alpha).
\end{eqnarray}
These equations are based on the approximation that the shear of the
outgoing null rays on the apparent horizon vanishes, which is true for
the case of a stationary black hole, and on the requirements that
$T_{\mu\nu}k^\mu k^\nu = T_{\mu\nu}\acute{k}^\mu\acute{k}^\nu =
T_{\mu\nu}k^\mu\acute{k}^\nu = 0$ on the apparent horizon if the black
hole is in quasi-equilibrium.  Finally, I reemphasize that the
conditions for quasi-equilibrium on the two expansions is given by
Eq.~(\ref{eq:QE-expans}), {\em not} by $\partial_t\sigma =
\partial_t\acute\sigma = 0$, so that the location of the horizon is
fixed but the black hole is allowed to rotate in ${\cal S}$.

On inspection, Eq.~(\ref{eq:QE-og-exp1}) yields the solution
\begin{equation}\label{eq:EQ-shift-cond}
	\beta_\perp = \alpha,
\end{equation}
which is compatible with the Killing horizon boundary condition
obtained in the case of corotation (\ref{eq:corot-shift-bc}).  The
Killing horizon boundary conditions for irrotation and general
rotation, (\ref{eq:irrot-shift-bc-1}) and
(\ref{eq:arb-rot-shift-bc-1}), are not compatible with
(\ref{eq:EQ-shift-cond}) unless we restrict the rotational terms to
act {\em in} the surface ${\cal S}$.  Thus, the correct boundary
condition on the shift for an irrotational black hole should be
\begin{equation}\label{eq:irrot-shift-bc}
	\beta^i|_{\cal S} = \alpha s^i|_{\cal S}
		+ \Omega h^i_j\left.\left(
		\frac\partial{\partial\phi}\right)^{\!\!j}\right|_{\cal S},
\end{equation}
and the condition for general rotation should be
\begin{equation}\label{eq:arb-rot-shift-bc}
	\beta^i|_{\cal S} = \alpha s^i|_{\cal S}
		- h^i_j\left[\Omega_\chi\chi^j
		- \Omega\left(\frac\partial{\partial\phi}\right)^{\!\!j}
		\right]_{\cal S}.
\end{equation}
Notice that our assumption of quasi-equilibrium has lead to one
condition on the shift, and yet Eqns.~(\ref{eq:corot-shift-bc}),
(\ref{eq:irrot-shift-bc}) and (\ref{eq:arb-rot-shift-bc}) specify
three conditions on the shift.  This is appropriate since we have made
specific choices for the rotation of the black hole, namely corotation,
irrotation, or some general rotation specified by $\Omega_\chi$ and
$\chi^i$.

The condition that ${\cal S}$ be an apparent horizon,
(\ref{eq:AH-cond}), also yields a useful boundary condition if we work
in the conformal space.  The conformal transformation on
$\gamma_{ij}$ (\ref{eq:conformal-metric}) induces a natural
conformal weighting for $h_{ij}$ and for the unit normal to ${\cal
S}$,
\begin{eqnarray}
\label{eq:conformal-h}
	h_{ij} &\equiv& \psi^4 \tilde{h}_{ij}, \\
\label{eq:conformal-s}
	s^i &\equiv& \psi^{-2}\tilde{s}^i.
\end{eqnarray}
We find that Eq.~(\ref{eq:AH-cond}) reduces to
\begin{equation}\label{eq:AH-BC}
	\tilde{s}^k\tilde\nabla_k\ln\psi|_{\cal S} =
		-{\textstyle\frac14}(\tilde{h}^{ij} \tilde\nabla_i\tilde{s}_j
		- \psi^2 J)|_{\cal S},
\end{equation}
which is a non-linear boundary condition that incorporates both $\psi$
and its normal derivative.  While this boundary conditions looks
somewhat complicated, it reduces to the often used minimal surface
boundary condition whenever $J|_{\cal S}=0$, and it has previously
been used successfully to construction initial-data
sets\cite{thornburg87}.

We can now further simplify the remaining condition given in
Eq.~(\ref{eq:QE-ig-exp1}).  From Eqns.~(\ref{eq:null-expansion-def1}),
(\ref{eq:null-expansion-def2}), and (\ref{eq:AH-cond}), we obtain
\begin{equation}\label{eq:ig-exp-redux1}
	\acute\sigma = \sqrt{2}J.
\end{equation}
Together with (\ref{eq:EQ-shift-cond}), (\ref{eq:conformal-h}), and
(\ref{eq:conformal-s}), this reduces (\ref{eq:QE-ig-exp1}) to
\begin{equation}\label{eq:QE-ig-exp2}
	J \tilde{s}^i\tilde\nabla_i\alpha|_{\cal S} =
		- \psi^2(J^2 - JK + \tilde{\cal D})\alpha|_{\cal S},
\end{equation}
with
\begin{equation}\label{eq:D2-conf-def}
	\tilde{\cal D} \equiv \psi^{-4}[
		\tilde{h}^{ij}(\breve\nabla_i - J_i)(\breve\nabla_j - J_j)
		- {\textstyle\frac12}\breve{R} + 2\breve\nabla^2\ln\psi].
\end{equation}
Here, $\breve\nabla_i$ and $\breve{R}$ are the covariant derivative
and Ricci scalar associated with the conformal metric
$\tilde{h}_{ij}$.  Equation~(\ref{eq:QE-ig-exp2}) represents a
complicated mixed condition on the lapse that involves an elliptic
operator acting over the closed surface ${\cal S}$.  There is no
guarantee that (\ref{eq:D2-conf-def}) is invertible, however, if it
is, then (\ref{eq:QE-ig-exp2}) represents a viable boundary condition.

\section{Quasi-circular orbits for black-hole binaries}
\label{sec:quasi-circ-orbits}
The primary goal of this paper is to fully define a formalism for
constructing initial data sets representing astrophysically realistic
black-hole binaries in quasi-circular orbits.  As was shown in
\S~\ref{sec:conf-thin-sandw}, the thin-sandwich decomposition of the
constraint equations appears to be a natural choice to use for this
purpose.  In particular, the simple choice of $\tilde{u}^{ij}=0$ is
required by the assumption of quasi-equilibrium.  In
\S~\ref{sec:quasi-equil-bound}, boundary conditions compatible with
quasi-equilibrium were developed.  However, there are still several
freely specifiable quantities that we have not considered within the
context of quasi-equilibrium.  In particular, we must consider how the
initial temporal gauge choice evolves off of the initial slice via
$\tilde\alpha$.  We must decide what value to use for the orbital
angular velocity, $\Omega$, in setting the outer-boundary conditions
(\ref{eq:asympt-shift-bc}).  And finally, we must choose the conformal
3-geometry, $\tilde\gamma_{ij}$, and the initial temporal gauge via
$K$.

\subsection{The temporal gauge choice}
\label{sec:temp-gauge-choice}
In all previous work on quasi-equilibrium data for either neutron
stars or black holes, maximal slicing has been used.  With this
choice, $K=0$, and it is natural to use the ``maximal slicing
equation'' to fix the lapse.  The maximal slicing equation has been
used extensively in numerical evolutions and is simply a linear
second-order elliptic equation for the lapse derived from the
evolution equation for $K$ by setting $\partial_tK=0$
(cf.\ Ref.~\cite{smarryork78b}).  For a quasi-equilibrium situation,
it is natural to use the condition $\partial_tK=0$, but to generalize
it to any arbitrary value for $K$.  In this case, the equation that
results,
\begin{eqnarray}\label{eq:const-trace-K-eqn}
	\tilde\nabla^2(\alpha\psi) - \alpha\left[
		{\textstyle\frac18}\psi\tilde{R}
		+ {\textstyle\frac5{12}}\psi^5K^2
		+ {\textstyle\frac78}\psi^{-7}\!\tilde{A}_{ij}\tilde{A}^{ij}
		\right.& \\ \mbox{} \left.
		+ 2\pi\psi^5K(\rho + 2S)\right] =&
		\psi^5\beta^i\tilde\nabla_{\!i}K, \nonumber
\end{eqnarray}
is often called the ``constant trace-$K$ equation'' (constant refers
to the fact that K is constant in time, not constant in space).

For a quasi-equilibrium situation, we would like to have
$\partial_t\gamma_{ij}=0$ and $\partial_tK_{ij}=0$.  However, this
will only be obtainable in true equilibrium situations where we have
exact Killing vectors.  We can decompose these conditions to get
$\partial_t\tilde\gamma_{ij}=0$, $\partial_t\psi=0$,
$\partial_t\tilde{A}_{ij}=0$ and $\partial_tK=0$ as the possible
quasi-equilibrium conditions that we might apply.  Of these, only
$\partial_t\tilde\gamma_{ij}=\tilde{u}_{ij}$ is part of the freely
specifiable data of the thin-sandwich decomposition.  Given that we
must fix $\tilde\gamma_{ij}$ and that we choose $\tilde{u}_{ij}=0$ as
a quasi-equilibrium condition, it is not reasonable to assume that we
can also find $\partial_t\tilde{A}_{ij}=0$ except in true equilibrium
conditions.  In addition to two dynamical degrees of freedom,
$\tilde\gamma_{ij}$ also encodes the initial spatial gauge choice.
Similarly, $K$ encodes the initial temporal gauge choice and parity
with the spatial gauge suggests that we should let quasi-equilibrium
fix $\partial_tK=0$ and that it is not reasonable to expect
$\partial_t\psi=0$ except in true equilibrium conditions.

We can examine the condition $\partial_t\psi=0$ even further.  From
Eq.~(\ref{eq:TS-gdot-decomp}) we find that this condition is
equivalent to
\begin{equation}\label{eq:dtpsi-cond}
	\tilde\nabla_k\beta^k + 6\beta^k\tilde\nabla_k\ln\psi
		- \psi^6\tilde\alpha K = 0.
\end{equation}
We might consider using (\ref{eq:dtpsi-cond}) to fix $K$ instead of
choosing it freely.  However, if we use (\ref{eq:dtpsi-cond}) to
replace $K$ in either Eq.~(\ref{eq:TS-momentum-const}) or
(\ref{eq:conf-mom}) we find that the resulting elliptic operator is
non-invertible.  One might then hope that an iterative scheme could be
used to fix $K$ so that $\partial_t\psi=0$ is satisfied.  However,
numerical experiments with this approach have shown it to be
unstable\cite{pfeiffer-pc-2001}.

Given that the thin-sandwich decomposition requires that
$\tilde\alpha$ be fixed, it is fortunate that the condition of
quasi-equilibrium provides an elliptic equation that fixes this
quantity, just as the condition of quasi-equilibrium provides the
means of constructing boundary conditions for the constraints.
Furthermore, quasi-equilibrium also provides a boundary
condition on $\tilde\alpha$ for use in (\ref{eq:const-trace-K-eqn}).
This boundary condition is given by Eq.~(\ref{eq:QE-ig-exp2}) with
$\alpha$ replaced by $\tilde\alpha$ via (\ref{eq:conformal-lapse}).
At spatial infinity, we have that $\tilde\alpha=1$.

\subsection{Obtaining circular orbits}
\label{sec:obta-circ-orbits}
Consider solving the thin-sandwich equations,
(\ref{eq:conf-Hamiltonian-const-v2}) and (\ref{eq:TS-momentum-const}),
in conjunction with the constant trace-$K$ equation,
(\ref{eq:const-trace-K-eqn}).  Assume that we fix a conformal
3-geometry $\tilde\gamma_{ij}$ (not necessarily flat), that we fix $K$
(not necessarily a maximal slice), and that we choose
$\tilde{u}^{ij}=0$.  We will demand that the apparent horizons
associated with each hole in our black-hole binary system occur on
specified closed 2-surfaces in the given 3-geometry.  On these inner
boundaries, we will use the conditions given in
Eqns.~(\ref{eq:AH-BC}), (\ref{eq:QE-ig-exp2}), and
(\ref{eq:corot-shift-bc}) or (\ref{eq:irrot-shift-bc}) together with
the asymptotic conditions that $\psi=1$, $\tilde\alpha=1$, and
(\ref{eq:asympt-shift-bc}).  Assuming that the system is well-posed,
solving it will result in two one-parameter families of solutions, one
for the corotation condition (\ref{eq:corot-shift-bc}) and one for the
irrotational condition (\ref{eq:irrot-shift-bc}).  These families of
solutions are parameterized by the orbital angular velocity, $\Omega$.

As is pointed out in GGB1, it is unreasonable to assume that the
entire family of solutions will satisfy the physical condition of
quasi-equilibrium in which we are interested.  This situation is
similar to that encountered in searching for quasi-circular orbits via
the effective potential method\cite{cook94e}.  In fact, the effective
potential in that approach was chosen on physical ground to pick those
solutions which represented approximately circular orbits.  However,
that effective potential was necessarily somewhat {\em ad-hoc}.  In my
opinion, the most important contribution of GGB1 is their more
physically well-founded and covariant method for determining which
members of the family of solutions satisfy the condition of
quasi-equilibrium.

The condition for choosing appropriate values of $\Omega$ given in
GGB1,2 is justified largely in terms of its relation to a general
relativistic version of the virial theorem\cite{gourgoulhon94}.  The
condition is that $\Omega$ is chosen so that the standard ADM mass,
$E_{\textsc{\tiny ADM}}$, and the Komar(or KVM) mass\cite{komar59},
$M_{\text{\tiny K}}$, agree.  In my view, the justification for this
condition resonates most clearly when one considers that, in general,
$E_{\textsc{\tiny ADM}} \neq M_{\text{\tiny K}}$.  However, as was
shown by Beig\cite{beig78} (see also Ref.~\cite{ashtekar79}), these
two mass measures agree when the spacetime is stationary.  Since a
quasi-equilibrium solution is approximating a stationary spacetime, it
is natural to demand that $E_{\textsc{\tiny ADM}} = M_{\text{\tiny
K}}$, and in this case, we should obtain quasi-circular orbits.

Since, for a general gauge choice, $E_{\textsc{\tiny ADM}}$ is not
necessarily given by the monopole piece of the conformal factor, we
will use a general definition of the ADM
mass\cite{york79},
\begin{equation}\label{eq:ADM-mass-gen-def}
	E_{\textsc{\tiny ADM}} = \frac1{16\pi}\oint_\infty{
		\gamma^{ij}\bar\nabla_k({\cal G}^k_i - \delta^k_i {\cal G})
		{\rm d}^2S_j}.
\end{equation}
Here, ${\cal G}_{ij} \equiv \gamma_{ij} - f_{ij}$, $f_{ij}$ is the
flat metric to which $\gamma_{ij}$ asymptotes, and ${\rm d}^2S_i$ is
the covariant surface area element.  We note that in
(\ref{eq:ADM-mass-gen-def}), indices can be raised and lowered with
either $\gamma_{ij}$ or $f_{ij}$, the trace, ${\cal G},$ can be
obtained with either metric, and $\bar\nabla_i$ can be replaced with
the flat covariant derivative.  Similarly, a general definition of the
Komar mass can be written as\cite{gourgoulhon94}
\begin{equation}\label{eq:komar-mass-gen-def}
	M_{\text{\tiny K}} = \frac1{4\pi}\oint_\infty{\gamma^{ij}
		(\bar\nabla_i\alpha - \beta^kK_{ik}){\rm d}^2S_j}.
\end{equation}
In many cases, $\beta^kK_{ik}$ will fall off faster than
$O(r^{-2})$ and the second term in (\ref{eq:komar-mass-gen-def})
will not contribute.  However, in some gauges this term is important.
An example is the Painlev\'e--Gullstrand coordinate system (cf.\
Ref.~\cite{kidder-etal-2000}).

\subsection{Quasi-equilibrium evolutionary sequences}
\label{sec:quasi-equil-evol}
It is highly desirable to be able to construct evolutionary sequences
of quasi-equilibrium binary configurations.  In particular, this
facilitates locating the ISCO.  When one considers neutron-star
configurations, there is a natural way to connect neighboring
solutions.  Namely, the number of baryons contained in each star
should not change as the star secularly evolves to smaller separation.
With black holes, there is no such conserved quantity and one must fix
some normalizing condition.

In order to construct evolutionary sequences from quasi-circular
orbits obtained by the effective-potential method, both the
normalizing condition and the orbital angular velocity had to be
determined by some means.  One condition that should be applicable to
evolutionary sequences of quasi-equilibrium
configurations\cite{ostriker-gun-69,hartle-70,Damour-etal-2000} is
\begin{equation}\label{eq:equilib-dEdj}
	\Omega = \frac{dE_{\textsc{\tiny ADM}}}{dJ}.
\end{equation}
This relation has been shown to hold well for evolutionary sequences
of both corotating\cite{baumgarte_etal98b} and
irrotational\cite{uryu-etal-2000} neutron-star binaries.  Because the
effective-potential method cannot determine the orbital angular
velocity of each configuration of holes in circular orbit,
Eq.~(\ref{eq:equilib-dEdj}) must be used to determine $\Omega$.  But,
locating each circular-orbit configuration, as well as determining
$\Omega$ requires a normalizing condition.  In work so far with the
effective-potential method, this normalization condition has been
chosen to be an {\em ad hoc} definition for the mass of an individual
black hole in the binary system, with the further assumption that this
mass remains constant as the black holes secularly evolve closer
together.  While this definition had appropriate limiting behavior
when the black holes have large separation, this normalization
condition seems problematical in the strong-field limit.

However, because quasi-equilibrium data constructed from the combination
of the thin-sandwich and constant trace-$K$ equations naturally yields
a value for $\Omega$ for each circular-orbit configuration, GGB2 point
out that (\ref{eq:equilib-dEdj}) can be used to set the normalization
condition for constructing evolutionary sequences.  This appears to be
a much more well-founded condition than the {\em ad hoc} condition
proposed in Ref.~\cite{cook94e} for the effective-potential method.
That {\em ad hoc} condition is essentially rooted in the notion that
the area of the apparent horizon, an estimate of the true irreducible
mass of a black hole, should remain constant during secular evolution.
Interestingly, the first results found in GGB2 show agreement with
the {\em ad hoc} method to the level of numerical error in the 
solutions.  Given the problems with the solutions obtained in GGB2,
as outlined in \S~\ref{sec:ggb-approach}, it is too soon to make
any conclusions regarding the validity of the {\em ad hoc} condition,
but this point should be considered further.

\section{Discussion}
\label{sec:discussion}
So far, we have not considered specific choices for the conformal
3-geometry, $\tilde\gamma_{ij}$, and initial slicing as specified by
$K$.  To date, only flat conformal 3-geometries and maximal slicings
(I will refer to these as the CFMS assumptions) have been explored
extensively.  It has been pointed out that quasi-equilibrium binary
systems in circular orbits will not exhibit a conformally flat
3-geometry at second post-Newtonian order\cite{rieth-Math-Grav}.  Some
consider this approximation to be the major source of error in current
numerical work to model quasi-equilibrium binary configurations and in
locating the ISCO\cite{Damour-etal-2000}.  Certainly, the assumption
of conformal flatness introduces {\em some} error.  But, it is not
clear that this is the most significant source of error in these
calculations.  As stated earlier, I suspect that the {\it ad hoc}
normalization condition used to determine $\Omega$ and construct
evolutionary sequences is the dominant source of error.  It will be
important to explore the sensitivity of solutions to the choices for
$\tilde\gamma_{ij}$ and $K$ in order to understand this issue better.

Of course, it is possible that no regular quasi-equilibrium solutions
exist when the CFMS assumptions are made.  As discussed in
\S~\ref{sec:ggb-approach}, a regular solution for the binary problem
does not seem possible when inversion-symmetric boundary conditions
are chosen becasue $\alpha|_{\cal S}=0$.  If we consider the case of a
single black hole, we can easily find a CFMS solution of
Eqns.~(\ref{eq:conf-Hamiltonian-const-v2}),
(\ref{eq:TS-momentum-const}), and (\ref{eq:const-trace-K-eqn}) that
satisfies the inner boundary conditions given by
Eqns.~(\ref{eq:EQ-shift-cond}), (\ref{eq:AH-BC}), and
(\ref{eq:QE-ig-exp2}).  This solution is simply the Schwarzschild
solution in isotropic coordinates.  In this case, although $\alpha=0$
on the apparent horizon, the solution is regular.  This is possible
because the spacetime admits a true Killing vector.  The direct
generalization of this solution to a binary system yields the problem
attempted in GGB1,2 that does not yield a regular solution.  However,
this does not mean that a quasi-equilibrium solution is impossible
given the CFMS assumptions.  The nonlinearity of the system of
equations and boundary conditions allow for at least the {\em
possibility} of such solutions, even if it seems unlikely.

The apparent problems with constructing a quasi-equilibrium, binary,
CFMS solution should not be taken to suggest that conformally flat
quasi-equilibrium binary solutions are unlikely.  The problem with
the CFMS assumptions is with maximal slicing, not conformal
flatness.  Again, for the case of a single black hole, we can
easily find solutions of the quasi-equilibrium equations and
boundary conditions that are conformally flat.  One example
is obtained from the Schwarzschild solution in Kerr--Schild
coordinates, also referred to as ingoing Eddington--Finkelstein
coordinates (cf.\ Ref.\cite{kidder-etal-2000}).  With the
radial coordinate transformation
\begin{equation}\label{eq:KS-isotropic-coord-trans}
	r = \frac{\tilde{r}}4 \left(1 + \sqrt{1+\frac{2M}{\tilde{r}}}\right)^2
		e^{2\left(1 - \sqrt{1 + \frac{2M}{\tilde{r}}}\right)},
\end{equation}
where $\tilde{r}$ is the usual areal radial coordinate in Kerr--Schild
coordinates, the spatial metric on the Kerr-Schild slicing is seen to
be conformally flat.  Equations~(\ref{eq:conf-Hamiltonian-const-v2}),
(\ref{eq:TS-momentum-const}), and (\ref{eq:const-trace-K-eqn}) are
obviously satisfied, and it is easy to verify that the
quasi-equilibrium boundary conditions given by
Eqns.~(\ref{eq:EQ-shift-cond}), (\ref{eq:AH-BC}), and
(\ref{eq:QE-ig-exp2}) are also satisfied on the horizon at
$\tilde{r}=2M$.  Of course, the slicing in this solution is not
maximal, or even asymptotically maximal.  A feature of this and other
similar non-maximal single hole solutions is that $\alpha\neq0$ on the
horizon.  It seems likely that this feature will also hold for similar
binary configurations, removing the most obvious obstacle confronting
the construction of regular quasi-equilibrium solutions.

It is clear that we will want to explore configurations that are not
constructed on a maximal slice and we will want to consider conformal
3-geometries that are not flat.  This idea has been explored in
similar contexts\cite{Kerrbinary98,Kerrbinary2000}, but little has
been done other than to demonstrate that solutions can be found.  The
major limitation of these works is that the boundary conditions used
were not well motivated.  However, some of the work done in
constructing appropriate values for $\tilde\gamma_{ij}$ and $K$ based
on the Schwarzschild and Kerr geometries should prove useful.  In the
context of quasi-equilibrium solutions, one should construct conformal
3-geometries and associated values of $K$ that approximate black hole
binaries that are at {\em rest} in the corotating coordinate system.
By superposing two Schwarzschild holes, one would approximate the
3-geometry of a pair of corotating black holes.  A superposition of
two Kerr holes could approximate the case of irrotational black holes,
assuming the spins are chosen correctly.  Again, it will be
interesting to asses the sensitivity of corotating and irrotational
solutions to the choice of the conformal 3-geometry.

An improvement in the choice for $\tilde\gamma_{ij}$ and $K$ would
come from taking these data from post-Newtonian solutions for binaries
in circular-orbits.  An appropriate metric could be constructed by
omitting radiation damping terms.  One possibility would be to revert
the effective one-body metric obtained in
Ref.~\cite{Buonanno-Damour-99} to an appropriate two body coordinate
system, assuming this can be done.  A metric accurate to 2.5PN order,
including spin effects, has been given explicitly in
Ref.~\cite{Tagoshi-etal-2001}.  Another possibility is found in
Ref.~\cite{Alvi-2000} where a post-Newtonian metric is matched to two
perturbed Schwarzschild metircs.  However, it is unclear to me whether
or not any of these solutions are immediately applicable.  A usable
metric must be written in terms of a gauge where the lapse and
3-metric {\em smoothly} cross the individual black-hole horizons.  In
particular, the lapse should not vanish at these horizons.

Finally, we might consider the use of a slicing and 3-metric obtained
from a post-Newtonian solution where radiation reaction {\em is}
included.  In this case, we could also obtain a non-zero solution for
$\tilde{u}^{ij}$.  However, we should be cautious in exploring
solutions with $\tilde{u}^{ij}\neq0$ since this violates the principle
of quasi-equilibrium.  In particular, the condition used to determine
the orbital angular velocity, $\Omega$, will most likely {\em not} be
applicable in this case.

\acknowledgments I would like to thank Mark Scheel and Harald
Pfeiffer for helpful discussions.  This work was supported in part by
NSF grant PHY-9988581.


\begin{thebibliography}{48}
\expandafter\ifx\csname natexlab\endcsname\relax\def\natexlab#1{#1}\fi
\expandafter\ifx\csname bibnamefont\endcsname\relax
  \def\bibnamefont#1{#1}\fi
\expandafter\ifx\csname bibfnamefont\endcsname\relax
  \def\bibfnamefont#1{#1}\fi
\expandafter\ifx\csname citenamefont\endcsname\relax
  \def\citenamefont#1{#1}\fi
\expandafter\ifx\csname url\endcsname\relax
  \def\url#1{\texttt{#1}}\fi
\expandafter\ifx\csname urlprefix\endcsname\relax\def\urlprefix{URL }\fi
\providecommand{\bibinfo}[2]{#2}
\providecommand{\eprint}[2][]{\url{#2}}

\bibitem[{\citenamefont{Cook et~al.}(1993)\citenamefont{Cook, Choptuik, Dubal,
  Klasky, Matzner, and Oliveira}}]{cook93}
\bibinfo{author}{\bibfnamefont{G.~B.} \bibnamefont{Cook}},
  \bibinfo{author}{\bibfnamefont{M.~W.} \bibnamefont{Choptuik}},
  \bibinfo{author}{\bibfnamefont{M.~R.} \bibnamefont{Dubal}},
  \bibinfo{author}{\bibfnamefont{S.}~\bibnamefont{Klasky}},
  \bibinfo{author}{\bibfnamefont{R.~A.} \bibnamefont{Matzner}},
  \bibnamefont{and} \bibinfo{author}{\bibfnamefont{S.~R.}
  \bibnamefont{Oliveira}}, \bibinfo{journal}{Phys. Rev. D}
  \textbf{\bibinfo{volume}{47}}, \bibinfo{pages}{1471} (\bibinfo{year}{1993}).

\bibitem[{\citenamefont{Brandt and Br\"{u}gmann}(1997)}]{brandt_bruegmann97}
\bibinfo{author}{\bibfnamefont{S.}~\bibnamefont{Brandt}} \bibnamefont{and}
  \bibinfo{author}{\bibfnamefont{B.}~\bibnamefont{Br\"{u}gmann}},
  \bibinfo{journal}{Phys. Rev. Lett.} \textbf{\bibinfo{volume}{78}},
  \bibinfo{pages}{3606} (\bibinfo{year}{1997}).

\bibitem[{\citenamefont{Cook}(1994{\natexlab{a}})}]{cook94e}
\bibinfo{author}{\bibfnamefont{G.~B.} \bibnamefont{Cook}},
  \bibinfo{journal}{Phys. Rev. D} \textbf{\bibinfo{volume}{50}},
  \bibinfo{pages}{5025} (\bibinfo{year}{1994}{\natexlab{a}}).

\bibitem[{\citenamefont{Baumgarte}(2000)}]{baumgarte-2000}
\bibinfo{author}{\bibfnamefont{T.~W.} \bibnamefont{Baumgarte}},
  \bibinfo{journal}{Phys. Rev. D} \textbf{\bibinfo{volume}{62}},
  \bibinfo{pages}{024018} (\bibinfo{year}{2000}).

\bibitem[{\citenamefont{Pfeiffer et~al.}(2000)\citenamefont{Pfeiffer,
  Teukolsky, and Cook}}]{pfeiffer-etal-2000}
\bibinfo{author}{\bibfnamefont{H.~P.} \bibnamefont{Pfeiffer}},
  \bibinfo{author}{\bibfnamefont{S.~A.} \bibnamefont{Teukolsky}},
  \bibnamefont{and} \bibinfo{author}{\bibfnamefont{G.~B.} \bibnamefont{Cook}},
  \bibinfo{journal}{Phys. Rev. D} \textbf{\bibinfo{volume}{62}},
  \bibinfo{pages}{104018} (\bibinfo{year}{2000}).

\bibitem[{\citenamefont{Kidder et~al.}(1992)\citenamefont{Kidder, Will, and
  Wiseman}}]{kidder92}
\bibinfo{author}{\bibfnamefont{L.~E.} \bibnamefont{Kidder}},
  \bibinfo{author}{\bibfnamefont{C.~M.} \bibnamefont{Will}}, \bibnamefont{and}
  \bibinfo{author}{\bibfnamefont{A.~G.} \bibnamefont{Wiseman}},
  \bibinfo{journal}{Class. Quantum Gravit.} \textbf{\bibinfo{volume}{9}},
  \bibinfo{pages}{L125} (\bibinfo{year}{1992}).

\bibitem[{\citenamefont{Damour et~al.}(2000)\citenamefont{Damour, Jaranowski,
  and Schafer}}]{Damour-etal-2000}
\bibinfo{author}{\bibfnamefont{T.}~\bibnamefont{Damour}},
  \bibinfo{author}{\bibfnamefont{P.}~\bibnamefont{Jaranowski}},
  \bibnamefont{and} \bibinfo{author}{\bibfnamefont{G.}~\bibnamefont{Schafer}},
  \bibinfo{journal}{Phys. Rev. D} \textbf{\bibinfo{volume}{62}},
  \bibinfo{pages}{084011} (\bibinfo{year}{2000}).

\bibitem[{\citenamefont{Bowen and York}(1980)}]{bowenyork80}
\bibinfo{author}{\bibfnamefont{J.~M.} \bibnamefont{Bowen}} \bibnamefont{and}
  \bibinfo{author}{\bibfnamefont{J.~W.} \bibnamefont{York},
  \bibfnamefont{Jr.}}, \bibinfo{journal}{Phys. Rev. D}
  \textbf{\bibinfo{volume}{21}}, \bibinfo{pages}{2047} (\bibinfo{year}{1980}).

\bibitem[{\citenamefont{Kulkarni et~al.}(1983)\citenamefont{Kulkarni, Shepley,
  and York}}]{ksy83}
\bibinfo{author}{\bibfnamefont{A.~D.} \bibnamefont{Kulkarni}},
  \bibinfo{author}{\bibfnamefont{L.~C.} \bibnamefont{Shepley}},
  \bibnamefont{and} \bibinfo{author}{\bibfnamefont{J.~W.} \bibnamefont{York},
  \bibfnamefont{Jr.}}, \bibinfo{journal}{Phys. Lett.}
  \textbf{\bibinfo{volume}{96A}}, \bibinfo{pages}{228} (\bibinfo{year}{1983}).

\bibitem[{\citenamefont{Matzner et~al.}(1999)\citenamefont{Matzner, Huq, and
  Shoemaker}}]{Kerrbinary98}
\bibinfo{author}{\bibfnamefont{R.~A.} \bibnamefont{Matzner}},
  \bibinfo{author}{\bibfnamefont{M.~F.} \bibnamefont{Huq}}, \bibnamefont{and}
  \bibinfo{author}{\bibfnamefont{D.}~\bibnamefont{Shoemaker}},
  \bibinfo{journal}{Phys. Rev. D} \textbf{\bibinfo{volume}{59}},
  \bibinfo{pages}{024015} (\bibinfo{year}{1999}).

\bibitem[{\citenamefont{Marronetti and Matzner}(2000)}]{Kerrbinary2000}
\bibinfo{author}{\bibfnamefont{P.}~\bibnamefont{Marronetti}} \bibnamefont{and}
  \bibinfo{author}{\bibfnamefont{R.~A.} \bibnamefont{Matzner}},
  \bibinfo{journal}{Phys. Rev. Lett.} \textbf{\bibinfo{volume}{85}},
  \bibinfo{pages}{5500} (\bibinfo{year}{2000}).

\bibitem[{\citenamefont{Cook et~al.}(1996)\citenamefont{Cook, Shapiro, and
  Teukolsky}}]{cook96b}
\bibinfo{author}{\bibfnamefont{G.~B.} \bibnamefont{Cook}},
  \bibinfo{author}{\bibfnamefont{S.~L.} \bibnamefont{Shapiro}},
  \bibnamefont{and} \bibinfo{author}{\bibfnamefont{S.~A.}
  \bibnamefont{Teukolsky}}, \bibinfo{journal}{Phys. Rev. D}
  \textbf{\bibinfo{volume}{53}}, \bibinfo{pages}{5533} (\bibinfo{year}{1996}).

\bibitem[{\citenamefont{Baumgarte et~al.}(1997)\citenamefont{Baumgarte, Cook,
  Scheel, Shapiro, and Teukolsky}}]{baumgarte_etal97b}
\bibinfo{author}{\bibfnamefont{T.~W.} \bibnamefont{Baumgarte}},
  \bibinfo{author}{\bibfnamefont{G.~B.} \bibnamefont{Cook}},
  \bibinfo{author}{\bibfnamefont{M.~A.} \bibnamefont{Scheel}},
  \bibinfo{author}{\bibfnamefont{S.~L.} \bibnamefont{Shapiro}},
  \bibnamefont{and} \bibinfo{author}{\bibfnamefont{S.~A.}
  \bibnamefont{Teukolsky}}, \bibinfo{journal}{Phys. Rev. Lett.}
  \textbf{\bibinfo{volume}{79}}, \bibinfo{pages}{1182} (\bibinfo{year}{1997}).

\bibitem[{\citenamefont{Baumgarte
  et~al.}(1998{\natexlab{a}})\citenamefont{Baumgarte, Cook, Scheel, Shapiro,
  and Teukolsky}}]{baumgarte_etal98b}
\bibinfo{author}{\bibfnamefont{T.~W.} \bibnamefont{Baumgarte}},
  \bibinfo{author}{\bibfnamefont{G.~B.} \bibnamefont{Cook}},
  \bibinfo{author}{\bibfnamefont{M.~A.} \bibnamefont{Scheel}},
  \bibinfo{author}{\bibfnamefont{S.~L.} \bibnamefont{Shapiro}},
  \bibnamefont{and} \bibinfo{author}{\bibfnamefont{S.~A.}
  \bibnamefont{Teukolsky}}, \bibinfo{journal}{Phys. Rev. D}
  \textbf{\bibinfo{volume}{57}}, \bibinfo{pages}{7299}
  (\bibinfo{year}{1998}{\natexlab{a}}).

\bibitem[{\citenamefont{Baumgarte
  et~al.}(1998{\natexlab{b}})\citenamefont{Baumgarte, Cook, Scheel, Shapiro,
  and Teukolsky}}]{baumgarte_etal98a}
\bibinfo{author}{\bibfnamefont{T.~W.} \bibnamefont{Baumgarte}},
  \bibinfo{author}{\bibfnamefont{G.~B.} \bibnamefont{Cook}},
  \bibinfo{author}{\bibfnamefont{M.~A.} \bibnamefont{Scheel}},
  \bibinfo{author}{\bibfnamefont{S.~L.} \bibnamefont{Shapiro}},
  \bibnamefont{and} \bibinfo{author}{\bibfnamefont{S.~A.}
  \bibnamefont{Teukolsky}}, \bibinfo{journal}{Phys. Rev. D}
  \textbf{\bibinfo{volume}{57}}, \bibinfo{pages}{6181}
  (\bibinfo{year}{1998}{\natexlab{b}}).

\bibitem[{\citenamefont{Marronetti et~al.}(1998)\citenamefont{Marronetti,
  Mathews, and Wilson}}]{marronetti-etal-1998}
\bibinfo{author}{\bibfnamefont{P.}~\bibnamefont{Marronetti}},
  \bibinfo{author}{\bibfnamefont{G.~J.} \bibnamefont{Mathews}},
  \bibnamefont{and} \bibinfo{author}{\bibfnamefont{J.~R.}
  \bibnamefont{Wilson}}, \bibinfo{journal}{Phys. Rev. D}
  \textbf{\bibinfo{volume}{58}}, \bibinfo{pages}{107503}
  (\bibinfo{year}{1998}).

\bibitem[{\citenamefont{Bonazzola et~al.}(1999)\citenamefont{Bonazzola,
  Gourgoulhon, and Marck}}]{bonazzola-etal-1999}
\bibinfo{author}{\bibfnamefont{S.}~\bibnamefont{Bonazzola}},
  \bibinfo{author}{\bibfnamefont{E.}~\bibnamefont{Gourgoulhon}},
  \bibnamefont{and} \bibinfo{author}{\bibfnamefont{J.-A.} \bibnamefont{Marck}},
  \bibinfo{journal}{Phys. Rev. Lett.} \textbf{\bibinfo{volume}{82}},
  \bibinfo{pages}{892} (\bibinfo{year}{1999}).

\bibitem[{\citenamefont{Marronetti et~al.}(1999)\citenamefont{Marronetti,
  Mathews, and Wilson}}]{marronetti-etal-1999}
\bibinfo{author}{\bibfnamefont{P.}~\bibnamefont{Marronetti}},
  \bibinfo{author}{\bibfnamefont{G.~J.} \bibnamefont{Mathews}},
  \bibnamefont{and} \bibinfo{author}{\bibfnamefont{J.~R.}
  \bibnamefont{Wilson}}, \bibinfo{journal}{Phys. Rev. D}
  \textbf{\bibinfo{volume}{60}}, \bibinfo{pages}{087301}
  (\bibinfo{year}{1999}).

\bibitem[{\citenamefont{Ury{\=u} and Eriguchi}(2000)}]{uryu-eriguchi-2000}
\bibinfo{author}{\bibfnamefont{K.}~\bibnamefont{Ury{\=u}}} \bibnamefont{and}
  \bibinfo{author}{\bibfnamefont{Y.}~\bibnamefont{Eriguchi}},
  \bibinfo{journal}{Phys. Rev. D} \textbf{\bibinfo{volume}{61}},
  \bibinfo{pages}{124023} (\bibinfo{year}{2000}).

\bibitem[{\citenamefont{Ury{\=u} et~al.}(2000)\citenamefont{Ury{\=u}, Shibata,
  and Eriguchi}}]{uryu-etal-2000}
\bibinfo{author}{\bibfnamefont{K.}~\bibnamefont{Ury{\=u}}},
  \bibinfo{author}{\bibfnamefont{M.}~\bibnamefont{Shibata}}, \bibnamefont{and}
  \bibinfo{author}{\bibfnamefont{Y.}~\bibnamefont{Eriguchi}},
  \bibinfo{journal}{Phys. Rev. D} \textbf{\bibinfo{volume}{62}},
  \bibinfo{pages}{104015} (\bibinfo{year}{2000}).

\bibitem[{\citenamefont{Gourgoulhon et~al.}(2001)\citenamefont{Gourgoulhon,
  Grandcl{\'e}ment, and Bonazzola}}]{gourgoulhon-etal-2001a}
\bibinfo{author}{\bibfnamefont{E.}~\bibnamefont{Gourgoulhon}},
  \bibinfo{author}{\bibfnamefont{P.}~\bibnamefont{Grandcl{\'e}ment}},
  \bibnamefont{and} \bibinfo{author}{\bibfnamefont{S.}~\bibnamefont{Bonazzola}}
  (\bibinfo{year}{2001}), \bibinfo{note}{preprint gr-qc/0106015}.

\bibitem[{\citenamefont{Grandcl{\'e}ment
  et~al.}(2001)\citenamefont{Grandcl{\'e}ment, Gourgoulhon, and
  Bonazzola}}]{gourgoulhon-etal-2001b}
\bibinfo{author}{\bibfnamefont{P.}~\bibnamefont{Grandcl{\'e}ment}},
  \bibinfo{author}{\bibfnamefont{E.}~\bibnamefont{Gourgoulhon}},
  \bibnamefont{and} \bibinfo{author}{\bibfnamefont{S.}~\bibnamefont{Bonazzola}}
  (\bibinfo{year}{2001}), \bibinfo{note}{preprint gr-qc/0106016}.

\bibitem[{\citenamefont{Cook}(1994{\natexlab{b}})}]{cook94f}
\bibinfo{author}{\bibfnamefont{G.~B.} \bibnamefont{Cook}}
  (\bibinfo{year}{1994}{\natexlab{b}}), \bibinfo{note}{unpublished notes}.

\bibitem[{\citenamefont{Lichnerowicz}(1944)}]{lichnerowicz-1944}
\bibinfo{author}{\bibfnamefont{A.}~\bibnamefont{Lichnerowicz}},
  \bibinfo{journal}{J. Math. Pures et Appl.} \textbf{\bibinfo{volume}{23}},
  \bibinfo{pages}{37} (\bibinfo{year}{1944}).

\bibitem[{\citenamefont{York}(1971)}]{york-1971}
\bibinfo{author}{\bibfnamefont{J.~W.} \bibnamefont{York}, \bibfnamefont{Jr.}},
  \bibinfo{journal}{Phys. Rev. Lett.} \textbf{\bibinfo{volume}{26}},
  \bibinfo{pages}{1656} (\bibinfo{year}{1971}).

\bibitem[{\citenamefont{York}(1972)}]{york-1972}
\bibinfo{author}{\bibfnamefont{J.~W.} \bibnamefont{York}, \bibfnamefont{Jr.}},
  \bibinfo{journal}{Phys. Rev. Lett.} \textbf{\bibinfo{volume}{28}},
  \bibinfo{pages}{1082} (\bibinfo{year}{1972}).

\bibitem[{\citenamefont{York}(1973)}]{york-1973}
\bibinfo{author}{\bibfnamefont{J.~W.} \bibnamefont{York}, \bibfnamefont{Jr.}},
  \bibinfo{journal}{J.~Math. Phys.} \textbf{\bibinfo{volume}{14}},
  \bibinfo{pages}{456} (\bibinfo{year}{1973}).

\bibitem[{\citenamefont{Bowen}(1979)}]{bowen79}
\bibinfo{author}{\bibfnamefont{J.~M.} \bibnamefont{Bowen}},
  \bibinfo{journal}{Gen. Relativ. Gravit.} \textbf{\bibinfo{volume}{11}},
  \bibinfo{pages}{227} (\bibinfo{year}{1979}).

\bibitem[{\citenamefont{York}(1999)}]{york-1999}
\bibinfo{author}{\bibfnamefont{J.~W.} \bibnamefont{York}, \bibfnamefont{Jr.}},
  \bibinfo{journal}{Phys. Rev. Lett.} \textbf{\bibinfo{volume}{82}},
  \bibinfo{pages}{1350} (\bibinfo{year}{1999}).

\bibitem[{\citenamefont{Wilson and Mathews}(1989)}]{Frontiers:Wilson}
\bibinfo{author}{\bibfnamefont{J.~R.} \bibnamefont{Wilson}} \bibnamefont{and}
  \bibinfo{author}{\bibfnamefont{G.~J.} \bibnamefont{Mathews}}, in
  \emph{\bibinfo{booktitle}{Frontiers in Numerical Relativity}}, edited by
  \bibinfo{editor}{\bibfnamefont{C.~R.} \bibnamefont{Evans}},
  \bibinfo{editor}{\bibfnamefont{L.~S.} \bibnamefont{Finn}}, \bibnamefont{and}
  \bibinfo{editor}{\bibfnamefont{D.~W.} \bibnamefont{Hobill}}
  (\bibinfo{publisher}{Cambridge University Press},
  \bibinfo{address}{Cambridge, England}, \bibinfo{year}{1989}), pp.
  \bibinfo{pages}{306--314}.

\bibitem[{\citenamefont{Bonazzola et~al.}(1997)\citenamefont{Bonazzola,
  Gourgoulhon, and Marck}}]{bonazzola97}
\bibinfo{author}{\bibfnamefont{S.}~\bibnamefont{Bonazzola}},
  \bibinfo{author}{\bibfnamefont{E.}~\bibnamefont{Gourgoulhon}},
  \bibnamefont{and} \bibinfo{author}{\bibfnamefont{J.-A.} \bibnamefont{Marck}},
  \bibinfo{journal}{Phys. Rev. D} \textbf{\bibinfo{volume}{56}},
  \bibinfo{pages}{7740} (\bibinfo{year}{1997}).

\bibitem[{\citenamefont{Cook and York}(1990)}]{cook90}
\bibinfo{author}{\bibfnamefont{G.~B.} \bibnamefont{Cook}} \bibnamefont{and}
  \bibinfo{author}{\bibfnamefont{J.~W.} \bibnamefont{York},
  \bibfnamefont{Jr.}}, \bibinfo{journal}{Phys. Rev. D}
  \textbf{\bibinfo{volume}{41}}, \bibinfo{pages}{1077} (\bibinfo{year}{1990}).

\bibitem[{\citenamefont{Thornburg}(1987)}]{thornburg87}
\bibinfo{author}{\bibfnamefont{J.}~\bibnamefont{Thornburg}},
  \bibinfo{journal}{Class. Quantum Gravit.} \textbf{\bibinfo{volume}{4}},
  \bibinfo{pages}{1119} (\bibinfo{year}{1987}).

\bibitem[{\citenamefont{Eardley}(1998)}]{eardley-1998}
\bibinfo{author}{\bibfnamefont{D.~M.} \bibnamefont{Eardley}},
  \bibinfo{journal}{Phys. Rev. D} \textbf{\bibinfo{volume}{57}},
  \bibinfo{pages}{2299} (\bibinfo{year}{1998}).

\bibitem[{\citenamefont{Smarr and York}(1978)}]{smarryork78b}
\bibinfo{author}{\bibfnamefont{L.} \bibnamefont{Smarr}} \bibnamefont{and}
  \bibinfo{author}{\bibfnamefont{J.~W.} \bibnamefont{York},
  \bibfnamefont{Jr.}}, \bibinfo{journal}{Phys. Rev. D}
  \textbf{\bibinfo{volume}{17}}, \bibinfo{pages}{1945} (\bibinfo{year}{1978}).

\bibitem[{\citenamefont{Pfeiffer}(2001)}]{pfeiffer-pc-2001}
\bibinfo{author}{\bibfnamefont{H.}~\bibnamefont{Pfeiffer}}
  (\bibinfo{year}{2001}), \bibinfo{note}{private communication}.

\bibitem[{\citenamefont{Gourgoulhon and Bonazzola}(1994)}]{gourgoulhon94}
\bibinfo{author}{\bibfnamefont{E.}~\bibnamefont{Gourgoulhon}} \bibnamefont{and}
  \bibinfo{author}{\bibfnamefont{S.}~\bibnamefont{Bonazzola}},
  \bibinfo{journal}{Class. Quantum Gravit.} \textbf{\bibinfo{volume}{11}},
  \bibinfo{pages}{443} (\bibinfo{year}{1994}).

\bibitem[{\citenamefont{Komar}(1959)}]{komar59}
\bibinfo{author}{\bibfnamefont{A.}~\bibnamefont{Komar}},
  \bibinfo{journal}{Phys. Rev.} \textbf{\bibinfo{volume}{113}},
  \bibinfo{pages}{934} (\bibinfo{year}{1959}).

\bibitem[{\citenamefont{Beig}(1978)}]{beig78}
\bibinfo{author}{\bibfnamefont{R.}~\bibnamefont{Beig}}, \bibinfo{journal}{Phys.
  Lett.} \textbf{\bibinfo{volume}{69A}}, \bibinfo{pages}{153}
  (\bibinfo{year}{1978}).

\bibitem[{\citenamefont{Ashtekar and Magnon-Ashtekar}(1979)}]{ashtekar79}
\bibinfo{author}{\bibfnamefont{A.}~\bibnamefont{Ashtekar}} \bibnamefont{and}
  \bibinfo{author}{\bibfnamefont{A.}~\bibnamefont{Magnon-Ashtekar}},
  \bibinfo{journal}{J.~Math. Phys.} \textbf{\bibinfo{volume}{20}},
  \bibinfo{pages}{793} (\bibinfo{year}{1979}).

\bibitem[{\citenamefont{York}(1979)}]{york79}
\bibinfo{author}{\bibfnamefont{J.~W.} \bibnamefont{York}, \bibfnamefont{Jr.}},
  in \emph{\bibinfo{booktitle}{Sources of Gravitational Radiation}}, edited by
  \bibinfo{editor}{\bibfnamefont{L.~L.} \bibnamefont{Smarr}}
  (\bibinfo{publisher}{Cambridge University Press},
  \bibinfo{address}{Cambridge, England}, \bibinfo{year}{1979}), pp.
  \bibinfo{pages}{83--126}.

\bibitem[{\citenamefont{Kidder et~al.}(2000)\citenamefont{Kidder, Scheel,
  Teukolsky, Carlson, and Cook}}]{kidder-etal-2000}
\bibinfo{author}{\bibfnamefont{L.~E.} \bibnamefont{Kidder}},
  \bibinfo{author}{\bibfnamefont{M.~A.} \bibnamefont{Scheel}},
  \bibinfo{author}{\bibfnamefont{S.~A.} \bibnamefont{Teukolsky}},
  \bibinfo{author}{\bibfnamefont{E.~D.} \bibnamefont{Carlson}},
  \bibnamefont{and} \bibinfo{author}{\bibfnamefont{G.~B.} \bibnamefont{Cook}},
  \bibinfo{journal}{Phys. Rev. D} \textbf{\bibinfo{volume}{62}},
  \bibinfo{pages}{084032} (\bibinfo{year}{2000}).

\bibitem[{\citenamefont{Ostriker and Gunn}(1969)}]{ostriker-gun-69}
\bibinfo{author}{\bibfnamefont{J.~P.} \bibnamefont{Ostriker}} \bibnamefont{and}
  \bibinfo{author}{\bibfnamefont{J.~E.} \bibnamefont{Gunn}},
  \bibinfo{journal}{Astrophys. J.} \textbf{\bibinfo{volume}{157}},
  \bibinfo{pages}{1395} (\bibinfo{year}{1969}).

\bibitem[{\citenamefont{Hartle}(1970)}]{hartle-70}
\bibinfo{author}{\bibfnamefont{J.~B.} \bibnamefont{Hartle}},
  \bibinfo{journal}{Astrophys. J.} \textbf{\bibinfo{volume}{161}},
  \bibinfo{pages}{111} (\bibinfo{year}{1970}).

\bibitem[{\citenamefont{Rieth}(1997)}]{rieth-Math-Grav}
\bibinfo{author}{\bibfnamefont{R.}~\bibnamefont{Rieth}}, in
  \emph{\bibinfo{booktitle}{Mathematics of Gravitation. {P}art {II}.
  {G}ravitational Wave Detection}}, edited by
  \bibinfo{editor}{\bibfnamefont{A.}~\bibnamefont{Kr{\'o}lak}}
  (\bibinfo{publisher}{Polish Academy of Sciences, Institute of Mathematics},
  \bibinfo{address}{Warsaw}, \bibinfo{year}{1997}), pp.
  \bibinfo{pages}{71--74}, \bibinfo{note}{proceedings of the Workshop on
  Mathematical Aspects of Theories of Gravitation held in Warsaw, February
  29--March 30, 1996}.

\bibitem[{\citenamefont{Buonanno and Damour}(1999)}]{Buonanno-Damour-99}
\bibinfo{author}{\bibfnamefont{A.}~\bibnamefont{Buonanno}} \bibnamefont{and}
  \bibinfo{author}{\bibfnamefont{T.}~\bibnamefont{Damour}},
  \bibinfo{journal}{Phys. Rev. D} \textbf{\bibinfo{volume}{59}},
  \bibinfo{pages}{084006} (\bibinfo{year}{1999}).

\bibitem[{\citenamefont{Tagoshi et~al.}(2001)\citenamefont{Tagoshi, Ohashi, and
  Owen}}]{Tagoshi-etal-2001}
\bibinfo{author}{\bibfnamefont{H.}~\bibnamefont{Tagoshi}},
  \bibinfo{author}{\bibfnamefont{A.}~\bibnamefont{Ohashi}}, \bibnamefont{and}
  \bibinfo{author}{\bibfnamefont{B.~J.} \bibnamefont{Owen}},
  \bibinfo{journal}{Phys. Rev. D} \textbf{\bibinfo{volume}{63}},
  \bibinfo{pages}{044006} (\bibinfo{year}{2001}).

\bibitem[{\citenamefont{Alvi}(2000)}]{Alvi-2000}
\bibinfo{author}{\bibfnamefont{K.}~\bibnamefont{Alvi}}, \bibinfo{journal}{Phys.
  Rev. D} \textbf{\bibinfo{volume}{61}}, \bibinfo{pages}{124013}
  (\bibinfo{year}{2000}).

\end{thebibliography}
\end{document}